\documentstyle[epsf]{ioplppt}
\eqnobysec
\begin{document}
\jl{3}
\title{Kondo disorder: a possible route towards non-Fermi liquid behavior}
\author{E Miranda\dag\ftnote{2}{To whom correspondence should be
addressed.}, V Dobrosavljevi\'{c}\dag\  and G Kotliar\S} 
\address{\dag\ National High Magnetic Field Laboratory, Florida State
University\\1800 E. Paul Dirac Dr., Tallahassee, Florida 32306.}
\address{\S\ Serin Physics Laboratory,
Rutgers University,
PO Box 849,
Piscataway NJ, 08855.}

\begin{abstract}

We present a general model of disorder in Kondo alloys that, under
certain conditions, leads to non-Fermi liquid behavior. The central
underlying idea is the presence of a distribution of local Kondo
temperature scales. If this distribution is broad enough, such that
there are sites with arbitrarily low Kondo temperatures, a non-Fermi
liquid phase is formed. We analyze thermodynamics and transport in this
approach and show it is consistent with a number of Kondo alloys. We
also compare the predictions of this model with the measured dynamical
magnetic response of these systems.

\end{abstract}


\section{Introduction}
\label{intro}

	Since its introduction by Landau in 1956, the Fermi liquid
paradigm has been the most robust stepping-stone of the theory of
metals\cite{landau}.  Its central assumption is the existence of a one
to one correspondence between the excitations of a free Fermi gas and an
interacting fermionic system. Even though originally formulated to
explain the behavior of a neutral Fermi system ($^3$He), it has formed
the basis of much of our understanding of what goes on in metallic
systems. It lies behind such successful theories as Migdal's
electron-phonon theory and provides the background upon which
instabilities such as superconductivity can be understood. It has proved
to be a valuable guide in the analysis of even very strongly correlated
systems, such as some of the heavy fermion metals, where the theory
seems to survive the most extreme circumstances, with enormous
renormalizations in the quasiparticle effective mass.

	The very idea of a correspondence between the low-lying
excitations of an interacting system and those of a reference one has
proved more general than the original strict application envisioned by
Landau. Indeed, this principle of ``adiabatic
continuity''\cite{anderson} can be found in very diverse situations with
very different and sometimes non-trivial reference systems. For example,
in the so-called Fermi liquid theory of the Kondo or Anderson impurity
problem, the reference system is a non-interacting scattering center
embedded in a free Fermi sea\cite{kondo_gen,localflt}. Other examples
include superfluid $^3$He\cite{he3legg}, atomic nuclei\cite{migdalnuc}
and interacting disordered metals\cite{castkotlee}. The corresponding
reference systems, in these latter cases, are the BCS model of pairing,
a shell model of the nucleus and a model of diffusive electrons in a
random potential, respectively.

	Given this operational definition of a Fermi liquid, the
question of whether a particular system behaves as a Fermi liquid or not
can be a rather delicate one. It is necessary first to determine what is
a natural reference system for a particular compound and then find out
whether the compound can fit a description which is adiabatically
connected to the reference one. We cannot overemphasize the fact that
this is not always a straightforward task.  Daunting as the task may be,
researchers have nonetheless developed rules of thumb to classify
various systems as non-Fermi liquids. One such commonly used criterion
is a resistivity which depends linearly on the temperature at low
temperatures, a result certainly inconsistent with a clean Fermi liquid.
However, as helpful as these empirical tests are, as a matter of
scientific rigor, we must stress that the classification of a certain
system as a non-Fermi liquid often involves careful cross-checking
between experiment and theory. This is particularly critical when the
system is disordered.  With these caveats in mind, we can proceed to
consider the question of non-Fermi liquid behavior.

	In recent years, a growing number of metallic compounds have
come to be known as counter-examples to the old Fermi liquid paradigm.
This is usually taken to mean that the adiabatic continuity hypothesis
appears to break down, {\em if one assumes the most natural reference
system for the system under study.} This has been perhaps most strongly
emphasized in the case of the high T$_c$ cuprate
superconductors\cite{hitc}, where Landau's original Fermi liquid theory
is unexpectedly violated in the normal state. In addition, some heavy
fermion systems have also been discovered which do not seem to fit the
general picture of a Fermi liquid\cite{uthpd2al3,maple}. The bulk of
this evidence has in effect turned the study of non-Fermi liquid
behavior into an independent frontier area of research whose
ramifications can only be dimly glimpsed. The main goal is to determine
{\em the possible routes} towards the breakdown of Fermi liquid theory
and to classify what new low-lying excitations are present in these
systems.

	Among the non-Fermi liquid candidates, some should be classified
as belonging to the heavy fermion family. This is taken to mean, in this
context, that for some range of composition or for certain values of
external parameters -- such as pressure or magnetic field -- these
compounds show typical heavy fermion physics with high temperature
incoherent Kondo behavior accompanied by the formation of a low
temperature heavy Fermi liquid, with or without magnetic or
superconducting order\cite{hf_review}. However, for the more interesting
values of these parameters, their behavior is not that of a Fermi
liquid.

	Some of the heavy fermion non-Fermi liquid metals have been
convincingly associated with the proximity to a quantum critical point.
These are the alloy CeCu$_{6-x}$Au$_x$\cite{nfl_cecuau} and the compound
CePd$_2$Si$_2$\cite{nfl_cepdsi}. Both systems show antiferromagnetism:
CeCu$_{6-x}$Au$_x$ for $x > 0.1$ and CePd$_2$Si$_2$ for pressures $P <
26$ {\rm kbar}. At the critical values of these parameters, the N\'eel
temperature $T_N$ vanishes and the behavior of the system appears to be
governed by the zero temperature critical point. The anomalous non-Fermi
liquid behavior of these compounds is exemplified by the resistivity
which is given by $\rho \approx \rho_0 + A T$ in CeCu$_{5.9}$Au$_{0.1}$
and as $\rho \approx \rho_0 + B T^{1.2}$ in CePd$_2$Si$_2$ at $P = 26$
{\rm kbar}. In the latter case, the non-Fermi liquid behavior is
interrupted by what appears to be a phase transition into a
superconducting state at $T_c \approx 0.4 K$.  The thermodynamic
response of CeCu$_{5.9}$Au$_{0.1}$ is also anomalous with the specific
heat $C/T \approx a {\rm ln}(T/T_0)$ and $\chi \approx
\chi_0 (1 - \alpha \sqrt{T})$. In addition,
external pressure can suppress $T_N$ to zero in CeCu$_{5.7}$Au$_{0.3}$
and the anomalous behavior associated with the quantum critical point
can be recovered. Though the non-Fermi liquid behavior can be reasonably
ascribed to the proximity of a quantum critical point a complete theory
does not yet seem to exist\cite{qcpth}.

	In contrast to the latter compounds, a series of other alloys
also seems to show characteristic non-Fermi liquid behavior which cannot
be clearly associated with quantum criticality. A partial list of these
is given in \Tref{tab1}, together with their corresponding resistivity,
specific heat and magnetic susceptibility (see also
Refs.~\cite{uthpd2al3,maple}). All of them show anomalous thermodynamic
and transport properties incompatible with a Fermi liquid description.

\begin{table}
\caption{Heavy fermion alloys which exhibit non-Fermi liquid behavior
not obviously ascribed to the proximity to a quantum critical point and
their properties. Here, $\rho(T)$ is the DC resistivity, $C(T)$ is the
specific heat, $\chi(T)$ is the magnetic susceptibility and
$1/\tau(\omega)$ is the frequency dependent scattering rate.  Below,
$A>0$ and $\omega_0>0$.\label{tab1}}\small
\begin{tabular}{@{}lllll}
\br
Compounds & $\rho(T)$ & $C(T)/T$ & $\chi(T)$ & $1/\tau(\omega) (T=0)$\\
\mr
UCu$_{5-x}$Pd$_x$\cite{ucupd,ucupdopt}& $ \rho_0 - A T$ & $ a
{\rm ln}(T_0/T)$ & $ a {\rm ln}(T_0/T)$ & $1/\tau_0 (1 - 
\omega/\omega_0)$ \\
M$_{1-x}$U$_x$Pd$_3$ (M=Sc,Y)\cite{mupd3,uypd3opt}& $ \rho_0 - A T$ & $
a {\rm ln}(T_0/T)$ & $ A T^{-0.3}$ & $1/\tau_0 (1 -
\omega/\omega_0)$ \\ 
La$_{1-x}$Ce$_x$Cu$_{2.2}$Si$_2$\cite{lacecu2si2} & $ \rho_0 - A T$ & $
a {\rm ln}(T_0/T)$ & $ a {\rm ln}(T_0/T)$ & ------ \\
U$_{1-x}$Th$_x$Pd$_2$Al$_3$\cite{uthpd2al3,uthpd2al3opt} & $ \rho_0 - A
T$ & $ a {\rm ln}(T_0/T)$ & $ \chi_0 - A \sqrt{T}$ & $1/\tau_0 (1 -
\omega/\omega_0)$ \\
Ce$_{1-x}$Th$_x$RhSb\cite{cethrhsb} & ------ & $
a {\rm ln}(T_0/T)$ & ------  & ------\\
U$_{x}$Th$_{1-x}$M$_2$Si$_2$ (M=Ru,Pt,Pd)\cite{uthm2si2} & $ \rho_0 + A
{\rm ln}T$ & $a {\rm ln}(T_0/T)$ & $a {\rm ln}(T_0/T) $ & ------\\
\br
\end{tabular}
\end{table}

	There are a few proposed scenarios to try to explain this
anomalous behavior. Some rely on the existence of a critical point at
$T=0$\cite{qcpth}. Though these theories should be relevant to the cases
of CeCu$_{6-x}$Au$_x$ and CePd$_2$Si$_2$ alluded to above, the alloys in
\Tref{tab1} are not obviously close to a phase boundary.

	On the other hand, other proposals have focused on a local
approach to the non-Fermi liquid physics. These include exotic impurity
models governed by a non-Fermi liquid fixed point such as the
quadrupolar Kondo model and various multi-channel Kondo
models\cite{exotimp,4chan}. These are largely based on the anomalous
behavior of a dilute system of such impurities. The inclusion of lattice
effects presents an additional challenge that has only recently started
to be addressed\cite{2chanlat}. Another proposed route to non-Fermi
liquid behavior has been to consider the competition between local
charge and spin fluctuations\cite{qimiao}. It is quite possible, perhaps
even likely, that the origin of the anomalous behavior is different for
different systems. We mention, in particular, the case of the last entry
of \Tref{tab1}, U$_{x}$Th$_{1-x}$M$_2$Si$_2$ (M=Ru,Pt,Pd), whose
description as a dilute system of magnetic two-channel
impurities\cite{exotimp} in Ref.~\cite{uthm2si2} appears to be fairly
good.

	The fact that the systems in \Tref{tab1} are all disordered
alloys immediately poses the question of the role of disorder in the
formation of the non-Fermi liquid state. Furthermore, recalling our
operational definition of a Fermi liquid, one must not neglect the fact
that the reference system in this case is almost certainly {\em a
disordered one.} The behavior of local moments in disordered systems has
been considered in past studies\cite{localdis1,localdis2,tkdist}.

	In this paper, we will focus on the first alloy of the table,
UCu$_{5-x}$Pd$_x$.  The accumulated experimental data on this system has
suggested to us that a model of disorder in f-electron systems is able
to account for its anomalous behavior. Paramount to this conclusion was
the Cu NMR study of Ref.~\cite{ucupdnmr}, which detected a large
inhomogeneous broadening of the NMR line, attributable to microscopic
disorder. It is the goal of this paper to present a scenario in which a
non-Fermi liquid state is generated as a consequence of the interplay of
disorder and strong correlations\cite{our}. This theory provides a
consistent way of understanding the non-Fermi liquid anomalies in both
transport and thermodynamic properties based on a single underlying
mechanism. The central idea of this theory is that moderate bare
disorder in a lattice model of localized moments is magnified due to the
strong local correlations between the moments and the conduction
electrons. In particular, a broad distribution of local energy scales
(Kondo temperatures) is generated\cite{tkdist}. A few local sites with
very small Kondo temperatures are {\em unquenched} at low temperatures
and dominate the thermodynamics and transport, forming a dilute gas of
low-lying excitations above the disordered metallic ground state.  The
presence of these {\em unquenched} moments leads to the formation of a
non-Fermi liquid phase. While we think that UCu$_{5-x}$Pd$_x$ ($x=1$ and
$x=1.5$) provides the best candidate system to exhibit such a phase to
this date, we expect this type of behavior to be seen in other Kondo
alloys. Whether the other alloys in \Tref{tab1} can be understood in
this framework is not clear at this moment.

	The outline of this paper is as follows. In \Sref{expbasis}, we
discuss the experimental work on the UCu$_{5-x}$Pd$_x$ alloys which
served as the motivation for this study, emphasizing the most important
underlying physical ideas. In \Sref{model}, we present the model and our
dynamical mean field theory approach to its solution. In \Sref{zerot},
we show the nature of the ground state of the model in the clean as well
as in the disordered limits. In \Sref{linearrho}, we explain the origin
of the linear dependence on the temperature of the resistivity. In
\Sref{dynsusc}, we apply this disorder model to the description of the
dynamical susceptibility and compare the results with experiments on
UCu$_{5-x}$Pd$_x$.  Finally, we conclude with a discussion of the
limitations of our approach in \Sref{discconc}. In order to lighten the
line of argument, a few derivations are left to the appendices.

\section{Experimental basis for the disorder model}
\label{expbasis}

	The major reason for considering disorder as the possible origin
of the anomalous behavior of the alloys of \Tref{tab1} was the Cu NMR
study reported in Ref.~\cite{ucupdnmr}. In this paper, the Cu NMR
field-swept powder pattern spectra of UCu$_{5-x}$Pd$_x$ for $x=1$ and
$x=1.5$ were shown to exhibit strong inhomogeneous broadening that could
only be explained by invoking the presence of short-range disorder.  A
simple disorder model was then used to describe both the broadening of
the NMR line and the spin susceptibility and specific heat measured in
these samples. The model consisted of a collection of independent spins,
mimicking the Uranium ions, each coupled to the conduction electron bath
by a dimensionless Kondo coupling constant $\lambda \equiv \rho_0 J$,
which was allowed to be randomly distributed in the samples. These were
supposed to mimic the local disorder induced by the $Pd$ substitution in
the $Cu$ ligand sites of the parent UCu$_5$ compound and, for
simplicity, were assumed to be distributed according to a Gaussian.
Assuming that the spins were completely uncorrelated, the thermodynamic
response was then calculated by taking an average over the response of a
single Kondo spin with the distribution of coupling constants. Since the
physics of a single Kondo spin is characterized by a single energy
scale\cite{kondo_gen} -- the Kondo temperature -- the important quantity
in the averaging procedure is the distribution of Kondo temperatures.

\begin{figure}
\epsfxsize=4.5in 
\centerline{\epsfbox{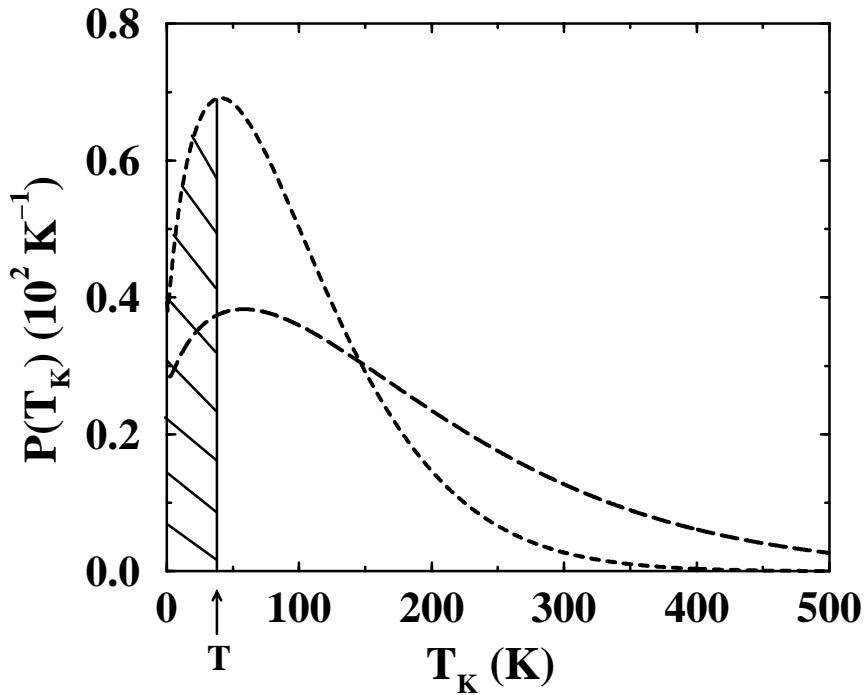}}
\caption{Experimentally determined distributions of
Kondo temperatures of the alloys UCu$_{5-x}$Pd$_x$ with
$x=1$(\longbroken) and $1.5$(\broken) (from
Ref.~\protect\cite{ucupdnmr}). The shaded area below $T$ represents the
low-$T_K$ spins which remain unquenched at that temperature. The upturn
at very low $T_K$'s is not shown (see text for discussion).
\label{fig1}}
\end{figure}

The Kondo temperature is defined as
\begin{equation}
T_K \equiv D e^{-1/\lambda}.
\label{kondotemp}
\end{equation}
Because of the exponential dependence on the coupling constant
$\lambda$, the corresponding distribution of Kondo temperatures is
skewed and considerably broadened. \Fref{fig1} shows the experimentally
determined distributions of Kondo temperatures for both alloys ($x=1$
and $x=1.5$).  We first note that, due to the Jacobian in the definition
of $P(T_K)$, it actually diverges weakly as $T_K \to 0$
\begin{equation}
P(T_K)=P[\lambda(T_K)] \left(\frac{d\lambda}{dT_K}\right) =
\frac{P[\lambda(T_K)]}{T_K \left({\rm ln}(D/T_K)\right)^2}.
\label{jacobian}
\end{equation}

However, for the distributions determined experimentally for
UCu$_{5-x}$Pd$_x$ ($x=1$ and $x=1.5$) and shown in \Fref{fig1}, the
point at which $P(T_K)$ shows an upturn and starts to diverge occurs at
very low temperatures: $T_K \approx 2.4 K$ ($x=1$) and $T_K \approx 0.8
K$ ($x=1.5$), which can hardly be distinguished on the scale of
\Fref{fig1}. None of the conclusions derived from the use of the full
$P(T_K)$ below depends on this small diverging tail, {\em as long as one
does not probe into the low temperature region below the upturn point
scale}.  Instead, as will be shown below, the response of the system is
dominated by the part of $P(T_K)$ {\em above the upturn}, which appears
to be tending to a constant value as $T_K \to 0$.  This means that there
is a finite number of spins with arbitrarily low $T_K$'s in the sample,
a feature that will be at the root of the non-Fermi liquid features. We
have, therefore, chosen to depict $P(T_K)$ only above the upturn point
and will confine the analysis below to a distribution which tends to a
constant as $T_K \to 0$. It remains an interesting possibility whether
the effects of the low $T_K$ divergence of $P(T_K)$ can be actually
observed. On the other hand, it is also conceivable that there is a
physical infrared cutoff $\lambda_c$ to $P(\lambda)$ ($\lambda_c
\stackrel{<}{\scriptstyle\sim} 0.11$ for both distributions). In the
absence of any experimental evidence for either possibility, we will
focus on the more relevant effect of the distribution of Kondo
temperatures above the upturn.

From the solution of the single impurity Kondo problem it is known that
thermodynamic quantities like the impurity spin susceptibility $\chi(T)$
increase with decreasing temperature in a Curie fashion, with
logarithmic corrections\cite{kondo_gen}. However, at very low
temperatures, the Curie-like divergence is cut off and the
susceptibility saturates to a constant value proportional to the inverse
Kondo temperature. The scale that separates the high temperature from
the low temperature region is the Kondo temperature. The impurity
specific heat divided by the temperature $C_V(T)/T$ has a somewhat
similar behavior. The saturation at the lowest temperatures is a
consequence of the ``disappearance'' of the free spin response, through
the formation of a singlet ground state with the conduction electron
bath, a process generally known as ``quenching''.

If we are measuring the thermodynamic properties at a certain finite
temperature $T$, there will always be Uranium ions with $T_K < T$, which
remain unquenched and whose contribution dominates the overall response
(the shaded area in \Fref{fig1}). As the temperature is lowered, the
number of such spins decreases, as more and more of them become
quenched. The thermodynamic behavior of the disordered system is,
therefore, dominated by the tails of the distribution of Kondo
temperatures, rather than by the average, a situation commonly known as
a Griffiths phase\cite{griffiths}. 

Since both $\chi(T)$ and $C_V(T)/T$ scale as the inverse Kondo
temperature at $T=0$, the fact that $P(T_K=0)
\not= 0$ immediately implies that the leading behavior of the averaged
quantity is a logarithmic divergence. Indeed, consider, for example, the 
susceptibility
\begin{equation}
\chi(T) \propto \frac{1}{T_K} f\left(\frac{T}{T_K}\right),
\label{susc}
\end{equation}
where the asymptotic forms of $f(x)$ are known\cite{kondo_gen}
\begin{equation}
f(x) \approx \left\{
\begin{array}{ll}
\alpha - \beta x^2 & x \ll 1; \\
\displaystyle
\frac{\gamma}{x} \left(1 - \frac{1}{{\rm ln}x} \right) & x \gg 1,
\end{array}
\right.
\label{chi_asymptot}
\end{equation}
$\alpha$, $\beta$ and $\gamma$ being universal numbers.  To find the
leading low temperature behavior it is sufficient to use the first term
in $P(T_K) = P_0 + P_1 T_K + \cdots$. Therefore, if $\langle
\cdots \rangle^{\rm av}$ denotes the average over the distribution of
Kondo temperatures
\begin{eqnarray}
\langle \chi(T) \rangle^{\rm av}& \propto \int_0^{\infty}
\frac{dT_K}{T_K} P(T_K) f\left(\frac{T}{T_K}\right) \nonumber \\
& \approx \int_0^{\Gamma/T} \frac{dy}{y} P_0 f\left(1/y\right),
\label{chi_aver}
\end{eqnarray}
where we cut off the integral by an arbitrary scale $\Gamma$ which sets
the region of validity of the approximate form for $P(T_K)$. From
\eref{chi_asymptot}, it is clear that the lower limit of the integral in
\eref{chi_aver} gives a regular contribution whereas the upper limit
dominates
\begin{equation}
\langle \chi(T) \rangle^{\rm av} \approx \int^{\Gamma/T} \frac{dy}{y}
\alpha P_0 \sim \alpha P_0 {\rm ln}\left(\frac{\Gamma}{T}\right).
\label{chi_log}
\end{equation}
A similar analysis holds for $C_V(T)/T$, with a similar divergence. 

	A caveat about the experimental situation is in order here. In
the literature, one often finds different power laws fitted to
thermodynamic and transport properties of this and other compounds. In
the particular case of UCu$_{5-x}$Pd$_x$, one can find $\chi(T) = \chi_0
T^{-\eta}$, with $\eta=0.27 \pm 0.03$\cite{ucupd},
$\eta=0.25$\cite{uthpd2al3} and $\eta=1/3$\cite{4chan,ucupddyn}. It is
not always clear what temperature range was used in the fits. In the
case of Ref.~\cite{4chan}, the $\eta=1/3$ power law is argued to be
valid in an intermediate range of temperatures (between $20$ and $300
K$). Besides, there is a small sample dependence to these quantities
and, in the case of the susceptibility, the magnetic field strength used
in the measurement can have an important effect\cite{ucupdnmr}. We feel
that all these aspects should be carefully considered when trying to
determine the temperature dependence.
 
	It is important to notice that the distribution of Kondo
temperatures reported in Ref.~\cite{ucupdnmr}, which is nearly
featureless at low $T_K$, gives a logarithmic divergence in $\chi(T)$
and $C_V(T)/T$ as the {\em leading} low temperature behavior. In the
fits shown in Ref.~\cite{ucupdnmr}, there are clear deviations from this
leading behavior at intermediate and higher temperatures. We point out
that power law behavior could be inferred from the analysis of a narrow
window of temperatures. What gives us particular confidence in the
disorder model is the fact that a full theoretical curve, with the added
complication of a finite magnetic field, can be well fitted to the
experiments. On the other hand, it is not clear how changes in the
specific form of the distribution function used would affect the quality
of the fit in the intermediate and high temperature regions.

\begin{figure}
\epsfxsize=4.5in 
\centerline{\epsfbox{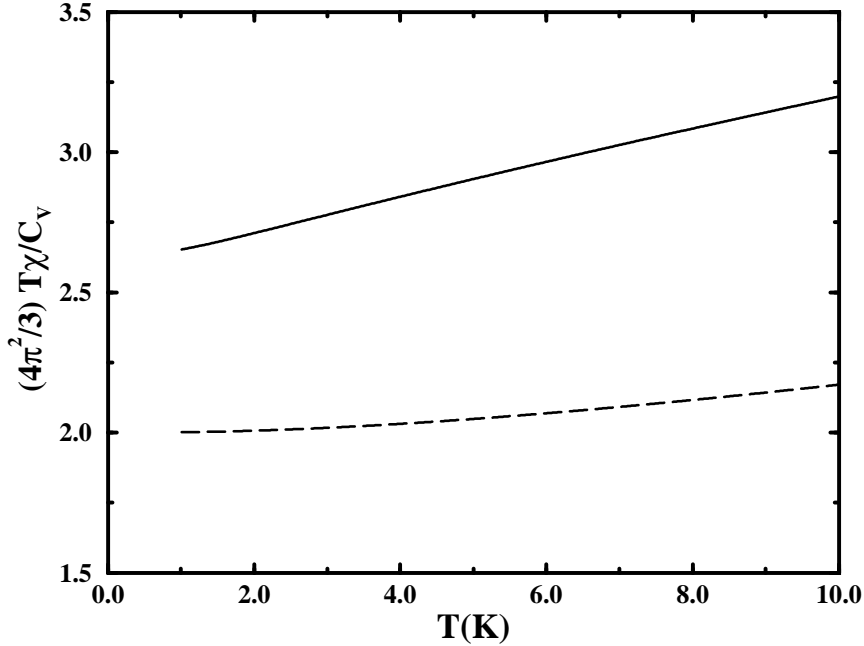}}
\caption{The Wilson ratio as a function of temperature as predicted by
the disorder model for the cases where $P(T_K=0)\not=0$ (\full) and
$P(T_K=0)=0$ (\broken). Note how the latter tends to the universal zero
temperature value of $2$.
\label{fig2}}
\end{figure}

	Finally, for completeness, we have plotted in \Fref{fig2} the
Wilson ratio ($R_W\equiv~T\chi(T)/C_V(T)$) prediction in the disorder
model as a function of temperature. When $P(0)=0$, the Wilson ratio
tends to its universal value of $2$ when $T \to 0$, which is consistent
with the Fermi liquid prediction. When $P(0)\not= 0$, the Wilson ratio
appears to tend to a different value for a wide range of temperatures.
This can be understood from \Eref{chi_log}. Because the scale $\Gamma$,
which enters this equation, depends on the details of the full scaling
curve, it is going to be, in general, different for $\chi(T)$ as
compared to $C_V(T)/T$. Though the scale is asymptotically irrelevant
when $T \gg \Gamma$, because of the slow logarithmic dependence, it
takes a temperature that is too low in order for one to be able to
observe the asymptotic behavior.

	We point out that one feature which distinguishes the alloy
UCu$_{5-x}$Pd$_x$ from the other alloys in \Tref{tab1} is the nature of
the doping. In the former case, the f-sublattice remains unchanged and
substitutions are introduced in the ligand sites. In the other cases,
the doping is performed directly in the f-sites. It is apparent that
ligand site substitution will affect the hybridization matrix element
$V$ of an Anderson model description and that will likely be the primary
effect (see \Sref{model} for a discussion of the disordered Anderson
lattice model). It is less certain what kind of effect such substitution
will have on the f-site energy $E_f$.  Moreover, lattice microstrains
should also be reflected in the hybridization amplitude $V$. This is
probably what happens in UCu$_4$Pd, which could be stoichiometric and
indeed appears to be so according the X-rays, but whose broad NMR lines
give unequivocal evidence of disorder. On the other hand, f-site
replacements can be mimicked by a different f-site energy $E_f$ ($E_f
\to + \infty$, when the substitutional ion is a non-magnetic ``Kondo
hole'') and, presumably, a different hybridization $V$. In the interest
of generality, and in the absence of a better understanding of the
nature of the disorder, we will consider in this paper {\em both} $E_f$
and $V$ disorder. What should be emphasized, however, is the fact that,
{\em whatever the microscopic origin of disorder}, its observable
consequences will always depend only on the distribution of Kondo
temperatures $P(T_K)$, for it is these scales which govern the measured
responses. In other words, almost all the information that can be
obtained experimentally about the disorder will be filtered through
$P(T_K)$.

	Therefore, the main conclusion of Ref.~\cite{ucupdnmr} was that
a simple model of Kondo disorder is able to account for the NMR line
broadening as well as the anomalous thermodynamic response of the alloys
studied.  This simple model relied on the assumption that the
correlations between the Kondo spins could be neglected, though they
form a concentrated lattice in these alloys. Past studies of heavy
fermion systems, both experimental\cite{doping} and
theoretical\cite{slavebos}, give us confidence that this may not be a
bad approximation for thermodynamic quantities (see also the
\ref{thermo}). However, it is well known that transport is {\em very}
different in dilute Kondo systems as opposed to concentrated Kondo
lattices. The so-called onset of coherence in the clean, concentrated
case is the most dramatic example. It is characterized by rather large
resistivities at high and intermediate temperatures, consequence of
strong incoherent Kondo scattering off the localized moments, which is
then followed by a precipitous fall by some orders of magnitude at the
lowest temperatures. At very low temperatures, the system can then be
characterized as a good metal. This low-temperature coherence is a
consequence of translational invariance and Bloch's theorem in the
ordered lattice system. We will, therefore, address the question of
whether a single unifying approach to the disorder problem, sensitive
enough to the formation of a coherent state in the clean limit, is able
to be formulated. We will answer the latter question in the affirmative
and will thereby show that the same mechanism that leads to diverging
thermodynamic properties at low $T$ -- the presence of spins with
arbitrarily low $T_K$'s -- will also predict a resistivity which is {\em
linear} in $T$, as observed in many non-Fermi liquid heavy fermion
alloys.

\section{The disordered Anderson lattice model and the dynamical mean
field theory equations}
\label{model}

	In order to consider the interplay of disorder and local moment
behavior we will focus our attention on a model of disordered Anderson
lattices. We will be guided by the series of alloys listed in
\Tref{tab1}. In the spirit of previous works on heavy fermion systems we
will assume that the f-electron sites can be described by simplified
non-degenerate Anderson impurities hybridized with a single broad
uncorrelated band of conduction electrons
\begin{equation}
\fl H = \sum\limits_{\vec{k}\sigma} \epsilon({\vec{k}})
c^{\dagger}_{\vec{k}\sigma} c^{\phantom{\dagger}}_{\vec{k}\sigma}
+ \sum\limits_{j\sigma} E^f_j f^{\dagger}_{j\sigma}
f^{\phantom{\dagger}}_{j\sigma} 
+ \sum\limits_{j\sigma} ( V_j c^{\dagger}_{j\sigma}
f^{\phantom{\dagger}}_{j\sigma}  + {\rm H. c.} )
+ U \sum\limits_{j} n_{fj\uparrow} n_{fj\downarrow},
\label{hammy}
\end{equation}
where, $c_{\vec{k}\sigma}$ destroys a conduction electron with momentum
$\vec{k}$ and spin $\sigma$ from a band with dispersion
$\epsilon({\vec{k}})$ and half bandwidth $D$ and $f_{j\sigma}$ destroys
an f-electron at site $j$ with spin $\sigma$.  Since $U$ is generally
the largest energy scale in typical f-shell parameters it will be taken
to infinity in this paper. 

The important thing to notice in \Eref{hammy} is that, unlike in the
usual periodic Anderson model, the local f-shell parameters $E^f_j$ and
$V_j$ are taken here to be random numbers distributed, in general,
according to two different distributions $P_1(E^f)$ and $P_2(V)$.  As
explained before, in the absence of a more detailed understanding of the
microscopic nature of the disorder in these systems, one should take
$P_1(E^f)$ and $P_2(V)$ to be, in principle, of the most general form.
Note, however, that we will {\em not} assume the disorder widths to be
too large. For instance, the NMR study of Ref.~\cite{ucupdnmr} suggests
bare distribution widths to be around 20\% of their average values.
However, as we will see, correlation effects will themselves generate
large {\em effective} disorder. In practice, we have considered Gaussian
and uniform distributions. Disorder in the conduction electron band,
though certainly present in these alloys, is not subject to these
correlation-induced renormalizations and has therefore been neglected in
our treatment. Despite these simplifications, we believe the model in
\eref{hammy} captures the essential ingredients relevant to the study of
disorder in Kondo alloys. 

Some previous studies of disordered Anderson or Kondo lattices have been
reported in the literature. We mention, in particular, the early work of
Te\v{s}anovi\'c, who employed a slave boson approach\cite{tesa}. More
recently, other studies of the effect of disorder in these model systems
have appeared\cite{infd_andlat_a,infd_andlat_b}.

In order to make progress, we have applied the dynamical mean field
theory of correlations and disorder to the Hamiltonian of
\Eref{hammy}\cite{infd_int,infd_dis,lisareview}. The derivation of the
dynamical mean field theory is most easily accomplished on a Bethe
lattice and we will focus on this class of models. In this case, the
conduction electron density of states acquires a simple semicircular
form. The solution of the full lattice problem then reduces to the
solution of an {\em ensemble} of impurity problems supplemented by a
self-consistency condition on the conduction electron Green's
function\cite{lisareview}. More precisely, the impurity problem action
for a random site $j$ is given, on the Matsubara frequency axis, by
\begin{equation}
S^{\rm imp}_j = T \sum_{\omega_n\sigma} \left[
f^{\dagger}_{j\sigma}(i\omega_n)
\left( - 
i\omega_n + E^f_j + V^2_j \Delta(i\omega_n) \right)
f^{\phantom{{\dagger}}}_{j\sigma}(i\omega_n) 
\right], 
\label{effaction}
\end{equation}
where the infinite-U constraint is implied and
\begin{equation}
\Delta(i\omega_n) = \frac{1}{i\omega_n + \mu - t^2
\overline{G}_c(i\omega_n)}.
\label{delta}
\end{equation}
Here, $t$ is the conduction electron hopping matrix element and
$\overline{G}_c(\omega)$ is the disorder-averaged local conduction
electron Green's function. Note that $\Delta(i\omega_n)$ is the
hybridization function of the conduction electrons which is ``seen'' by
each local f-site. Therefore, it corresponds to the local conduction
electron Green's function with one f-site removed. This is the so-called
``cavity'' Green's function\cite{lisareview}.

The self-consistency condition determines $\overline{G}_c(\omega)$
through 
\begin{equation}
\overline{G}_c(i\omega_n) = \left\langle \frac{1}{i\omega_n + \mu - t^2
\overline{G}_c(i\omega_n) - \Phi_j(i\omega_n)} \right\rangle^{\rm av},
\label{gc}
\end{equation}
where
\begin{equation}
\Phi_j(i\omega_n) =
\frac{ V^2_j}{i\omega_n - E^f_j - \Sigma^{\rm imp}_{fj}(i\omega_n)}.
\label{phi}
\end{equation}
Here, $\left\langle \dots \right\rangle^{\rm av}$ denotes the average
over disorder defined by the distribution functions $P_1(E^f)$ and
$P_2(V)$. $\Sigma^{\rm imp}_{fj}(i\omega_n)$ is the f-electron
self-energy derived from the impurity model of \eref{effaction}. This is
explicitly defined by
\begin{equation}
\Sigma^{\rm imp}_{fj}(i\omega_n) = i\omega_n - E^f_j -
V^2_j \Delta(i\omega_n) - G^{\rm imp}_{fj}(i\omega_n),
\label{fsig}
\end{equation}
where the f-electron Green's function is given by
\begin{eqnarray}
G^{\rm imp}_{fj}(\tau) &= - \langle T f^{\phantom{{\dagger}}}_j(\tau)
f^{\dagger}_{j}(0) \rangle^{\rm imp}_j \nonumber \\
G^{\rm imp}_{fj}(i\omega_n) &= \int^{\beta}_0 d\tau G^{\rm
imp}_{fj}(\tau) e^{i\omega_n\tau}.
\label{fgreen}
\end{eqnarray}
Here, we have used $\langle \dots \rangle^{\rm imp}_j$ to denote the
quantum mechanical/thermal average under the action of \Eref{effaction}.

Once the problem defined by \eref{effaction}--\eref{fgreen} has been
solved, the conduction electron self-energy $\Sigma_c(i\omega_n)$ is
obtained from 
\begin{equation}
\fl \overline{G}_c(i\omega_n) = \int_{-2t}^{2t} d\epsilon
\frac{\rho_0(\epsilon)}{i\omega_n + \mu - \epsilon - \Sigma_c(i\omega_n)};
\qquad \rho_0(\epsilon) = \frac{1}{\pi t}\sqrt{1 - (\epsilon/2t)^2}.
\label{sigmac}
\end{equation}
This self-energy is the important object for the calculation of the
conductivity, which, in the infinite coordination limit, involves no
vertex corrections\cite{infd_cond}
\begin{equation}
\fl \sigma(\omega) = \frac{(2te)^2}{\hbar \pi a}
\int_{-\infty}^{+\infty} d\nu 
\int_{-\infty}^{+\infty} d\epsilon \frac{f(\nu)-f(\nu+\omega)}{\omega}
\rho_0(\epsilon) A(\epsilon,\nu) A(\epsilon,\nu+\omega),
\label{cond}
\end{equation}
where $f(\nu)$ is the Fermi function and we have introduced a lattice
parameter $a$ and the correct dimensional factors appropriate for three
dimensions. In \eref{cond}, $A(\epsilon,\omega)$ is the conduction
electron one-particle spectral density
\begin{equation}
A(\epsilon,\omega) = {\rm Im} G_c(\epsilon,\omega) \equiv
{\rm Im} \left(\frac{1}{\omega + \mu -
\epsilon - \Sigma_c(\omega)}\right).
\label{specdens}
\end{equation}
The DC conductivity is then given by
\begin{equation}
\sigma_{DC} = \frac{(2te)^2}{\hbar \pi a} \int_{-\infty}^{+\infty} d\nu
\left( - \frac{\partial f}{\partial
\nu}\right) \int_{-\infty}^{+\infty} d\epsilon
\rho_0(\epsilon) A^2(\epsilon,\nu).
\label{dccond}
\end{equation}
This expression can be further simplified as shown in
\ref{app_expcond}.  To make contact with the usual Drude formula for
the DC conductivity it is useful to define the scattering time
\begin{equation}
\tau \equiv \frac{1}{\pi \rho_0(\mu)} \int_{-\infty}^{+\infty} d\nu
\left( - \frac{\partial f}{\partial
\nu}\right) \int_{-\infty}^{+\infty} d\epsilon
\rho_0(\epsilon) A^2(\epsilon,\nu),
\label{tau}
\end{equation}
which reduces to the usual case when $\tau \gg 1/D$
\begin{equation}
\tau \longrightarrow \frac{1}{2{\rm Im}\Sigma_c(0)} \qquad
(\tau \gg 1/D).
\label{tauXsigma}
\end{equation}
In \ref{app_cself_tmat}, a formula which expresses the conduction
electron self-energy in terms of the disorder-averaged impurity T-matrix
is derived. This formula will be useful in \Sref{linearrho}, when we
analyze the temperature dependence of the resistivity.

Thermodynamic properties within this approach are not drastically
different from the predictions of the simple disorder model of
Ref.~\cite{ucupdnmr}, discussed in \Sref{expbasis}. Though the
self-consistently determined conduction electron density of states
``seen'' by the f-sites ($\Delta(\omega)$) can be substantially modified
as compared to its free value, these modifications only lead to a
renormalized density of states, which is the quantity that ultimately
enters into the Kondo temperature expression. Once the renormalized
density of states is given, the prediction of the dynamical mean field
theory is essentially the same as that of the simple disorder model of
Ref.~\cite{ucupdnmr}, the difference being negligible. We discuss the
thermodynamics properties in the dynamical mean field theory in
\ref{thermo}.

We would like to emphasize at this point what processes are included and
what processes are left out of this approach. When the interaction $U$
is turned off, the treatment of the disorder problem that is obtained is
equivalent to the well-known coherent potential approximation
(CPA)\cite{cpa}. This approximation is known to give reliable results as
long as localization effects are negligible, since the conduction
electron density of states fluctuations are treated on the average. This
seems to be a safe approximation in the case of the alloys of interest,
where estimates based on the zero temperature DC resistivity give
$k_{F}l \approx 3-10$.  However, the interplay of disorder fluctuations
and correlations on the f-sites is fully kept in our treatment and,
indeed, is at heart of the physics that will emerge.  This is evidenced
by the fact that one needs to correctly solve an {\em ensemble} of
interacting impurity problems in order to close the equations.  Each
member of the {\em ensemble} of impurity problems has an associated
characteristic Kondo temperature $T_K$, thus generating a distribution
of local Kondo scales\cite{tkdist}. The fluctuations associated with
this distribution of Kondo temperatures will be responsible for the
anomalous low-temperature behavior in the non-Fermi liquid regime.
Other processes which are left out of this essentially local approach
are related to the RKKY interaction between f-sites mediated by the
exchange of conduction electron spin fluctuations. We will comment on
the possible limitations of this approximation in \Sref{discconc}.

\section{The zero temperature state}
\label{zerot}

The most difficult part of solving \eref{effaction}--\eref{fgreen} is
solving the impurity problem. Let us consider initially the clean case.
Then, there is only one impurity model to be solved. Equations \eref{gc}
and \eref{sigmac} are then trivially solved and yield $\Sigma_c(\omega)
= \Phi(\omega)$. At low temperatures, the impurity problem is governed
by a Fermi liquid fixed point\cite{kondo_gen,localflt} and $\Sigma^{\rm
imp}_f(\omega)$ can be parameterized at low temperatures and energies
as\cite{hewson_fl}
\begin{equation}
\Sigma^{\rm imp}_f(\omega) \approx a + b \omega +i c \left( \omega^2 +
\pi^2 T^2\right),
\label{fsig_fl}
\end{equation}
where, $a$, $b$ and $c$ are constants. From this, it follows that 
\begin{equation}
\Sigma_c(\omega) \approx \frac{zV^2}{\omega - \tilde{\epsilon}_f - i z c
\left( \omega^2 + \pi^2T^2 \right)},
\label{csig_fl}
\end{equation}
where we have redefined $a$ and $b$ in terms of a renormalized f-level
energy $\tilde{\epsilon}_f$ and a wave function renormalization factor
$z$. The first thing to notice in this expression is that at $T=0$,
$\Sigma_c(0)$ is real, leading to a purely real conduction electron
self-energy in the pure case. This is the hallmark of the low
temperature coherent transport of the clean system, a consequence of
translational invariance and a feature naturally incorporated in the
dynamical mean field theory. Furthermore, the Fermi liquid form of the
impurity problem self-energy ultimately leads to the Fermi liquid
behavior of the conduction electron lattice self-energy. In particular,
the resistivity derived from \eref{csig_fl} will exhibit a
characteristic $T^2$ law.

Let us now move on to the disordered case at zero temperature.  We have
employed the slave boson mean field theory to solve the {\em ensemble}
of impurity problems at $T=0$\cite{slavebos}. This theory is known to
provide a good description of the infinite-U Anderson impurity model at
temperatures $T \ll T_K$.  It will also serve as a good starting point
for the understanding of the complete solution of our dynamical mean
field equations at an arbitrary temperature. Details of the slave boson
treatment of the impurity problem are given in \ref{app_slbos}. 
These solutions to the {\em ensemble} of impurity problems were then
used in an iteration scheme to solve the full set
\eref{effaction}--\eref{fgreen}. 

\begin{figure}
\epsfxsize=5.in 
\centerline{\epsfbox{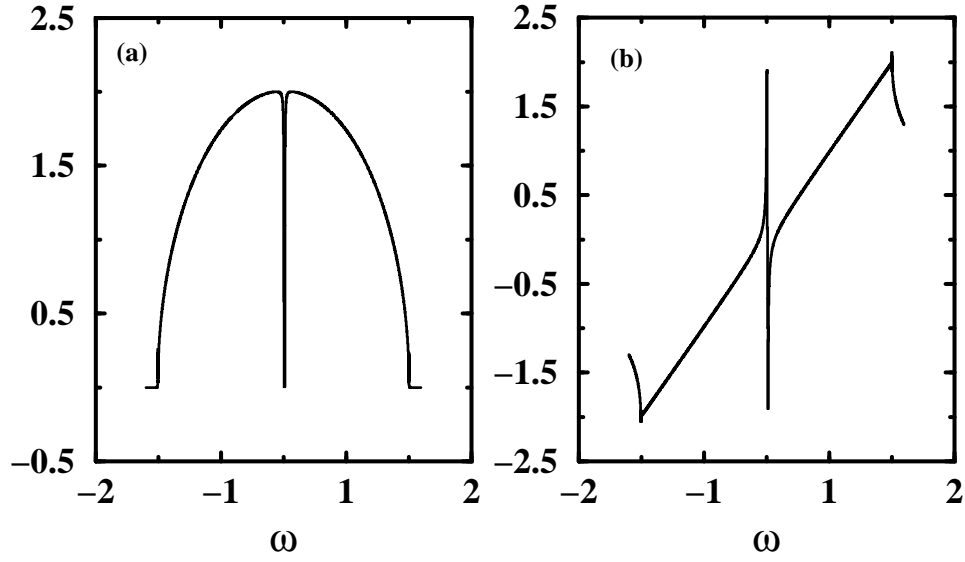}}
\caption{(a) Imaginary and (b) real part of the conduction electron
local Green's function as a function of frequency at $T=0$. The
parameters used were: $D=1$, $V=0.1$, $\mu=0$, $E^f = -0.05$ and there
is no disorder. 
\label{fig3}}
\end{figure}

\begin{figure}
\epsfxsize=5.in 
\centerline{\epsfbox{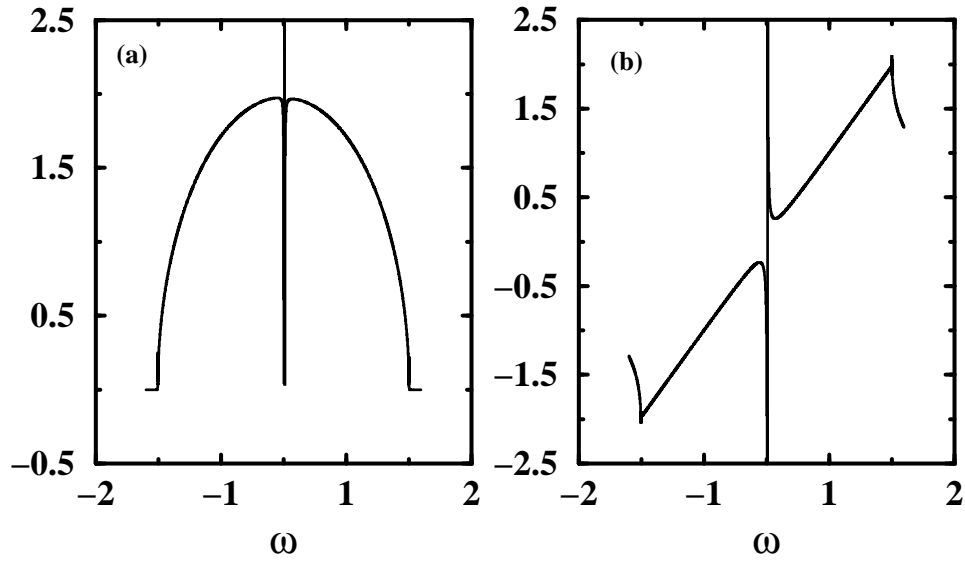}}
\caption{(a) Imaginary and (b) real part of the conduction electron
``cavity'' Green's function $\Delta(\omega)$ (see \protect\eref{delta})
as a function of frequency at $T=0$. The parameters used were: $D=1$,
$V=0.1$, $\mu=0$, $E^f = -0.05$ and there is no disorder.  
\label{fig4}}
\end{figure}

\Fref{fig3} shows the self-consistent solution to $G_c(\omega)$ in the
clean case and \Fref{fig4} shows the corresponding ``cavity'' Green's
function $\Delta(\omega)$. The only modification introduced by the
lattice of f-sites on the conduction electron density of states is the
appearance of a gap centered around the renormalized f-level position
$\epsilon_f$. This is familiar from the slave boson large-N solution of
the infinite-U Anderson lattice\cite{slavebos}. The scale of the gap is
given by $\approx r^2 V^2/D$, which is the Kondo temperature scale. The
important feature of the imaginary part of the ``cavity'' Green's
function $\Delta(\omega)$ is the appearance of a delta function at the
center of the gap $\epsilon_f$, which corresponds to the removal of one
f-site.

\begin{figure}
\epsfxsize=4.5in 
\centerline{\epsfbox{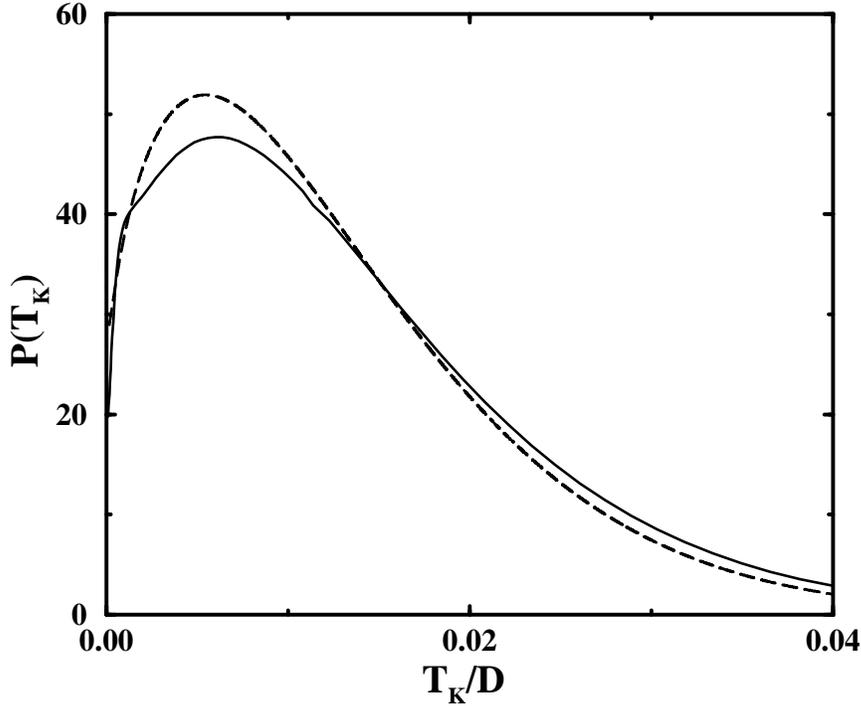}}
\caption{Typical distribution of Kondo temperatures obtained in the
fully self-consistent solution of the dynamical mean field theory
(\full) and the distribution appropriate for UCu$_4$Pd from
Ref.~\protect\cite{ucupdnmr}(\broken). The distribution was a Gaussian
and the parameters used were: $D=1$, $\mu=0$, $E^f=-1$, $\langle V^2
\rangle =0.17$, $W_{V^2} = 0.032$. The upturn at very low $T_K$'s is not
shown. 
\label{fig5}}
\end{figure}

The presence of disorder introduces important changes in these
structures at the scales given by the range of Kondo temperatures
produced as a consequence of the distribution of local f-parameters.
These are defined here as
\begin{equation}
T_{Kj} \equiv
\sqrt{(\epsilon^f_j-\tilde{\Delta}_j'(0))^2+(\tilde{\Delta}_j''(0))^2},
\label{tkdef}
\end{equation}
where $\tilde{\Delta}_j(\omega) \equiv r^2_j V^2_j \Delta(\omega)$ and
we are denoting real and imaginary parts by single and double primes,
respectively. In the Kondo limit, they acquire the familiar form
\begin{equation}
T_{Kj} \rightarrow D \exp{\left(E^f_j/(2\rho_0V_j^2)\right)},
\label{tkkonlim}
\end{equation}
where $\rho_0 = \Delta''(0)/\pi$. \Fref{fig5} shows a typical
distribution of Kondo temperatures from a fully self-consistent solution
of the zero temperature problem. For comparison, we have also plotted
the $P(T_K)$ used in the fit to the susceptibility of UCu$_4$Pd in
Ref.~\cite{ucupdnmr}. The self-consistent distribution has a structure
very similar to the experimentally determined one.

\begin{figure}
\epsfxsize=4.5in 
\centerline{\epsfbox{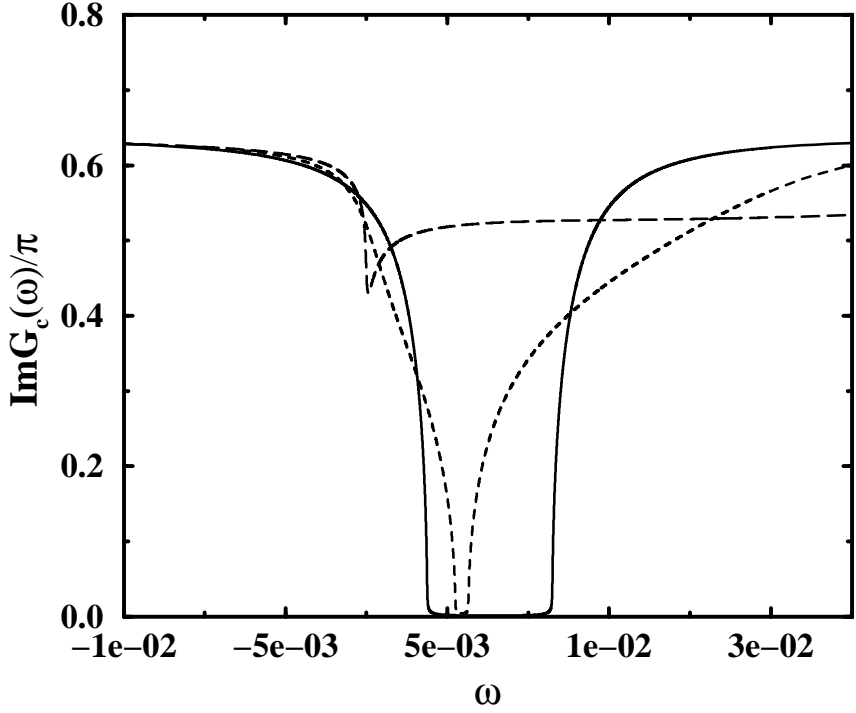}}
\caption{Imaginary part of the conduction electron Green's function
as a function of frequency at $T=0$ for different amounts of disorder in
the $E^f$ parameter. A uniform distribution was used with widths
$W=0.001$ (\full), $W=0.05$ (\broken) and $W=0.15$ (\longbroken). The
parameters used were: $D=1$, $V=0.1$, $\mu=0$, $\langle E^f \rangle =
-0.05$. 
\label{fig6}}
\end{figure}

\begin{figure}
\epsfxsize=4.25in 
\centerline{\epsfbox{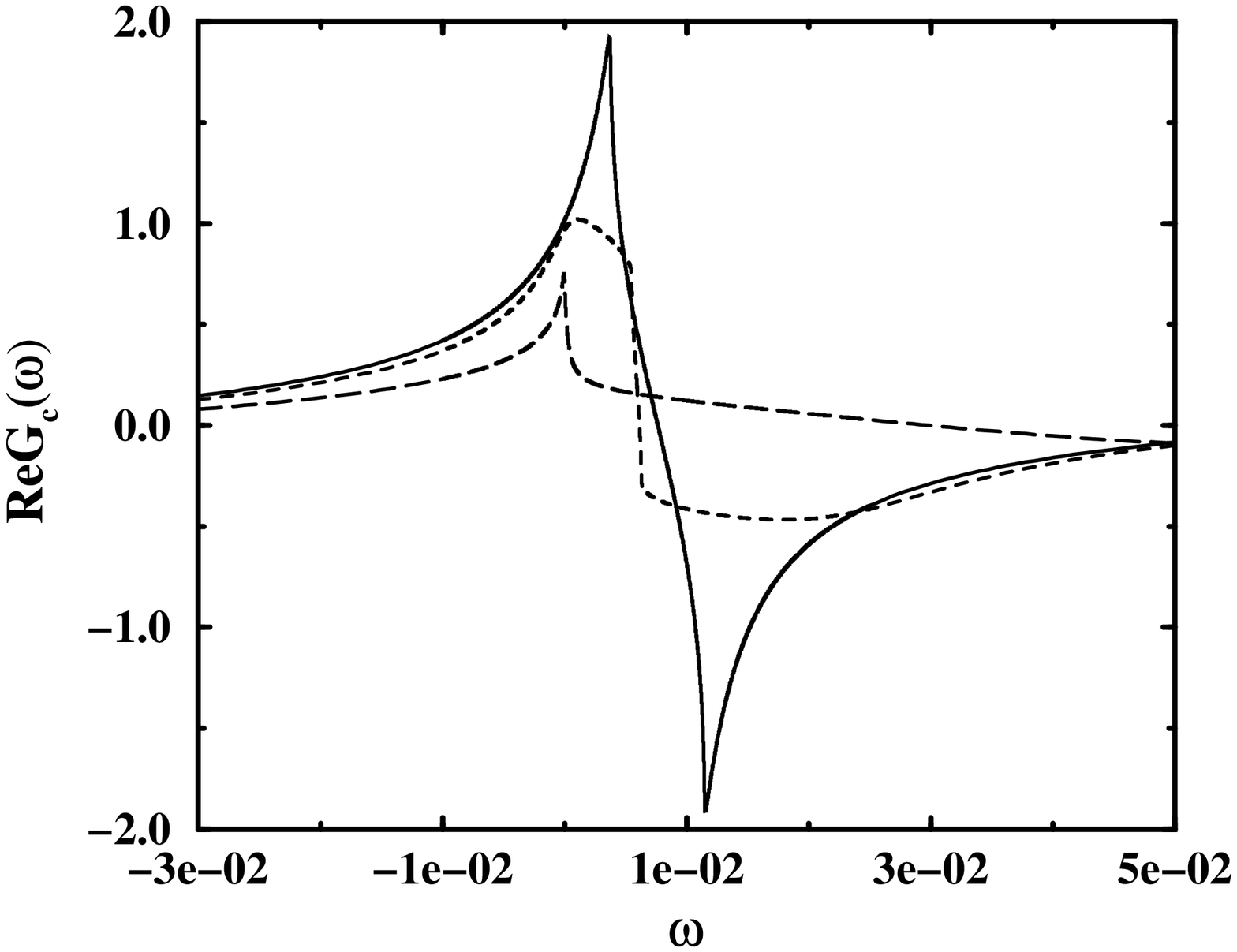}}
\caption{Real part of the conduction electron Green's function
as a function of frequency at $T=0$ for different amounts of disorder in
the $E^f$ parameter. A uniform distribution was used with widths
$W=0.001$ (\full), $W=0.05$ (\broken) and $W=0.15$ (\longbroken). The
same parameters were used as in \protect\Fref{fig3}. 
\label{fig7}}
\end{figure}

\begin{figure}
\epsfxsize=4.in 
\centerline{\epsfbox{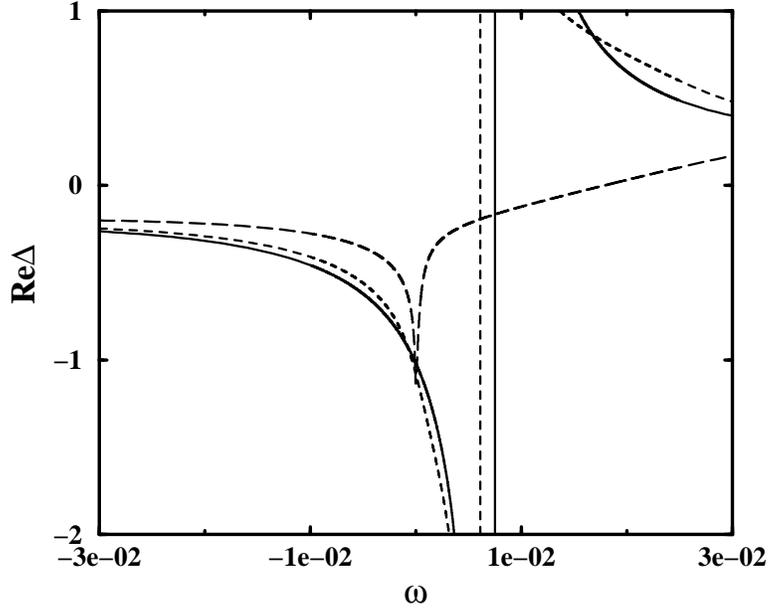}}
\caption{Real part of the conduction electron ``cavity'' Green's
function $\Delta(\omega)$ as a function of frequency at $T=0$ for
different amounts of disorder in the $E^f$ parameter. A uniform
distribution was used with widths $W=0.001$ (\full), $W=0.05$ (\broken)
and $W=0.15$ (\longbroken). The same parameters were used as in
\protect\Fref{fig3}. 
\label{fig8}}
\end{figure}

These modifications are illustrated in figures~\ref{fig6}, \ref{fig7}
and \ref{fig8}, which show the influence of various amounts of disorder
in the gap structure of $G_c(\omega)$ and $\Delta(\omega)$. Only the low
energy region is shown since that is the only region that is strongly
influenced by the introduction of disorder. It is clear from
\Fref{fig6}, that sufficient disorder can close the gap in the
conduction electron density of states. Qualitatively, a distribution of
Kondo energy scales for the {\em ensemble} of impurity problems will be
reflected in a distribution of energy gaps. Their sizes and positions
are essentially determined by the $T_{Kj}$'s and $\epsilon^f_j$'s,
respectively, and sufficient disorder in their distributions will lead
to an ultimate closing of the gap. In the metallic regime being
considered here, the disappearance of the gap does not lead to dramatic
effects, since the chemical potential lies away from the gap region.
However, there might be important consequences in the Kondo insulator or
Anderson insulator regimes. Corresponding modifications are also present
in the real parts of $G_c(\omega)$ and $\Delta(\omega)$, as seen in
figures \ref{fig7} and \ref{fig8}. The result is similar in the case
where $E^f$ is uniform and $V_j$ is distributed, since the relevant
scales are the Kondo temperatures \eref{tkkonlim}. However, due to the
different dependences of $T_{Kj}$ on $E^f_j$ and $V_j$, their
distributions will be quantitatively different and will reflect
differently on the $T_{Kj}$ distribution and consequently on the
transport properties.

The central aspect of the interplay between disorder and correlations is
an {\em enhancement} of the bare disorder by the local Kondo physics of
the impurity {\em ensemble}. This can be quite easily understood by an
examination of the qualitative aspects of the solution at $T=0$.  At
zero temperature, only $\Phi_j(0)$ enters into the calculation of
$\Sigma_c(0)$ or $G_c(0)$, either of which, in turn, is enough for the
calculation of the DC conductivity (see \eref{dccond} or
\eref{dccondsimpl}).  As was emphasized before, given a certain
distribution of the quantities $\Phi_j$, \Eref{gc} corresponds to a CPA
treatment of disorder in the conduction band, where the $\Phi_j$'s play
the role of scattering potential strengths. Interactions are important
in determining the value these latter quantities acquire, given a
self-averaged conduction electron bath through $\Delta(\omega)$.

From the Fermi liquid analysis, the zero temperature, zero energy form
of $\Phi_j(0)$ is real and given by
\begin{equation}
\Phi_j(0)= - \frac{V^2_j}{E^f_j + \Sigma^{\rm imp}_j(0)}.
\label{lowenphi}
\end{equation}
In the slave boson mean field theory, it is given by (see \eref{phi_sb}) 
\begin{equation}
\Phi_j(0)= - \frac{r^2_jV^2_j}{\epsilon^f_j}.
\label{lowenphi_sb}
\end{equation}
Now, in the Kondo limit $|E^f_j| >> \rho_0 V^2_j$ ($E^f_j < 0$), it is
easy to show from the mean field equations \eref{mfeqsa} and
\eref{mfeqsb} that 
\begin{equation}
\Phi_j(0) \longrightarrow \frac{1}{\Delta'(0)} \qquad \left(|E^f_j| >>
\rho_0 V^2_j, E^f_j < 0 \right).
\label{phi_kl}
\end{equation}

\begin{figure}
\epsfxsize=4.5in 
\centerline{\epsfbox{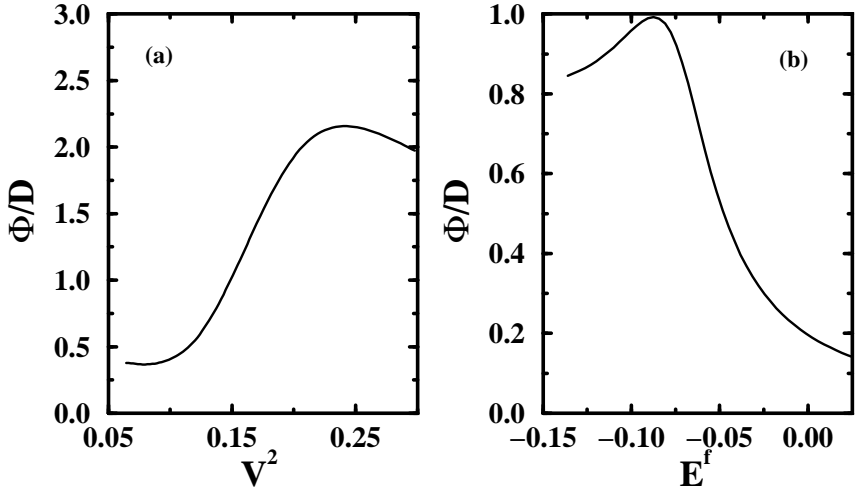}}
\caption{Variation of the effective local scattering potential strength
$\Phi(0)$ as a function of $V^2$ and $E^f$. Notice how $\Phi(0)$ varies
on the scale of conduction electron parameters ($D$) {\em not} on the
scale of local f-parameters. The distributions used were Gaussian and
the parameters were $D=1$, $\mu=0$, (a) $E^f=-1$, $\langle V^2 \rangle
=0.17$, $W_{V^2} = 0.032$, (b) $V=0.1$, $\langle E^f \rangle = -0.055$
and $W_{E^f}=0.027$.
\label{fig9}}
\end{figure}

The important point here is the fact that the impurity problem
parameters have disappeared from the local effective scattering
potential strength and its scale is now given by the conduction electron
band scale. As can be seen from \Fref{fig8}, the real part of the
``cavity'' Green's function that enters \eref{phi_kl} spans a wide range
of values in the region close to the chemical potential (here set to
zero). Because of the distribution of local parameters $\epsilon^f_j$
and $r_j$, this whole region close to $\mu$ will be probed by the
different f-sites and $\Phi_j(0)$ will also vary. However, its variation
will be on the conduction electron scale. Indeed, we have plotted in
\Fref{fig9} the variation of $\Phi_j(0)$ for a given strength of
disorder, in both cases of random $E^f_j$'s and random $V_j$'s. It is
clear that its distribution range is {\em not} on the same scale as the
variation of local f-parameters, which is very narrow, but rather, on
the scale of $D$.  {\em Therefore, the effective disorder seen by the
conduction electrons is considerably enhanced due to the local f-shell
correlation effects.}

\begin{figure}
\epsfxsize=4.5in 
\centerline{\epsfbox{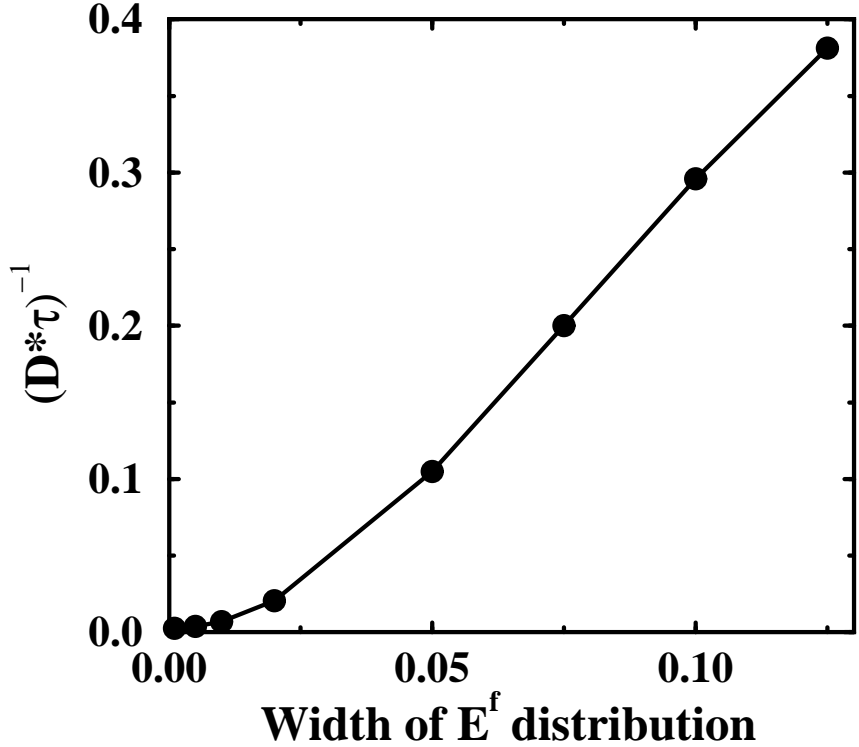}}
\caption{Scattering rate as a function of the width of the uniform $E^f$
distribution.  The strong correlations in the f-shell produce an
enhanced effective disorder. The parameters used were: $D=1$, $\mu=0$,
$V=0.1$, $\langle E^f \rangle = -0.05$.
\label{fig10}}
\end{figure}

As a result of this enhancement effect, large resistivities can be
generated even though the range of variation of the f-shell parameters
is very narrow. This is illustrated in \Fref{fig10}, where the
scattering rate $1/\tau$, as given from \eref{tau}, is plotted as a
function of the width of the uniform $E^f$ distribution. It is clear
that modest amounts of disorder in $E^f$ produce rather substantial
scattering rates. Thus, although the clean system exhibits coherent
transport at low $T$, {\em the introduction of a small amount of f-site
disorder leads to the destruction of these coherence effects.}
Qualitatively, once the lattice effects of coherence are destroyed by
sufficient disorder, the resistivity as a function of temperature will
then resemble the independent Kondo impurity results, with its
characteristic decreasing resistivity with increasing temperature. This
point was nicely illustrated in a recent study\cite{infd_andlat_b} of
binary alloy disorder in the Anderson lattice, which relies on a similar
treatment of correlations and disorder. Furthermore, several doping
studies of heavy fermion systems seem to bear out the above
picture\cite{doping}.

We summarize now the two major results of the study of disorder
discussed in this section:
\begin{itemize}
\item Due to the local Kondo physics at each f-site, the {\em
effective} disorder generated from a bare distribution of local f-shell
parameters is strongly renormalized up to scales of the order of the
conduction electron bandwidth.
\item Although the clean system has low resistivities due to the onset of
coherence at low $T$, moderate amounts of f-shell disorder are capable
of destroying this low-$T$ coherence, leading to characteristic incoherent
Kondo scattering behavior.
\end{itemize}

\section{The linear temperature dependence of the resistivity}
\label{linearrho}

Having established the incoherent nature of the transport with
sufficient disorder strength, we now focus on the temperature dependence
of the resistivity in this strongly correlated disordered state.  We
will rely on \ref{app_cself_tmat}, which relates the conduction electron
self-energy $\Sigma_c(\omega)$ to the averaged impurity model T-matrix
$T^{\rm imp}_j(\omega)$ (\Eref{sigmac3}). At sufficiently high
temperatures, compared to the highest $T_K$ in the distribution, the
averaged T-matrix at low energies becomes very small, reflecting the
weak coupling nature of the impurity response at high temperatures.
Therefore, there will be only a small contribution to the conduction
electron self-energy and consequently to the resistivity. At zero
temperature, as discussed in \Sref{zerot}, sufficient disorder will kill
all coherence in the transport properties and the system will exhibit a
rather large resistivity. What happens at low temperatures?

\begin{figure}
\epsfxsize=4.5in 
\centerline{\epsfbox{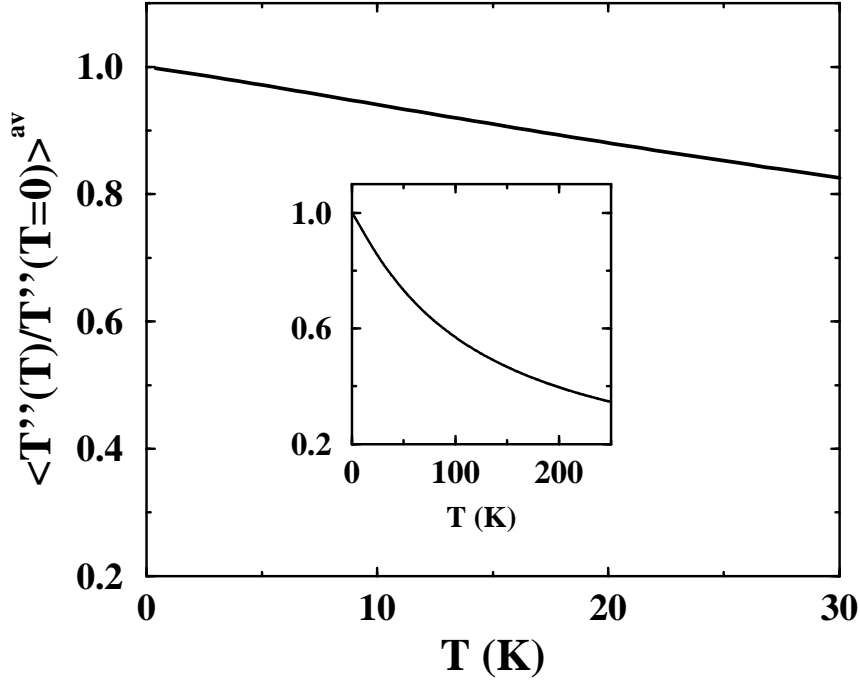}}
\caption{Temperature dependence of the imaginary part of the single
impurity T-matrix averaged over the disorder distribution appropriate
for UCu$_{3.5}$Pd$_{1.5}$, as determined experimentally in
Ref.~\protect\cite{ucupdnmr}. The inset shows the same quantity over a
wider temperature range.
\label{fig11}}
\end{figure}

For that, it is convenient to analyze the average impurity T-matrix. We
have therefore plotted in \Fref{fig11}, the imaginary part of the
impurity T-matrix, using the distribution of Kondo temperatures derived
from the fit to the thermodynamic properties of
UCu$_{5-x}$Pd$_x$\cite{ucupdnmr}. It is clear that the leading
temperature dependence is linear. This leads to a linear conduction
electron self-energy and to a linear resistivity through \Eref{dccond},
consistent with the behavior observed in several of the alloys in
\Tref{tab1}. At higher temperatures, there are clear deviations from the
leading linear behavior as can be seen in the inset. Can we understand
what conditions are necessary for this anomalous non-Fermi liquid
behavior?

The essential condition for the linear temperature dependence of the
resistivity, like the thermodynamic response discussed in
\Sref{expbasis}, is that the distribution of Kondo temperatures be such
that $P(T_K=0) \not= 0$. In other words, there must be a finite fraction
of f-sites with arbitrarily small Kondo temperatures. This can be made
more apparent by an analysis similar to the one in \Sref{expbasis}.
However, one has to be careful of how one understands the averaging
process. This is because, unlike the thermodynamic responses, transport
properties {\em cannot be averaged over single-impurity results.} This
is made evident in the analysis of the zero temperature state of
\Sref{zerot}. The single impurity result amounts to a strong, almost
unitary scattering center at $T=0$, with a correspondingly large
resistivity. Any attempt to simply average this result would lead to a
large resistivity. However, due to translational invariance, it is clear
that the clean sample can only have a vanishing resistivity, a result
which encounters a natural description in the dynamical mean field
theory, as was explained. Even in the disordered case, only the CPA-like
averaging process embodied in \Eref{gc} has any sense.  However, given a
disordered ground state {\em deviations from zero temperature} can still
be analyzed in a fashion that resembles the averaging of thermodynamic
quantities. This is because, at low temperatures, a few low-$T_K$ spins
are unquenched and cease to contribute significantly to the scattering.
Since they form a {\em dilute} system of subtracted scattering centers,
their contribution is additive and can therefore be averaged.

With these caveats in mind, we can proceed to analyze the conduction
electron self-energy. If the temperature is raised from $0$ to $T$,
there will be corresponding variations in all quantities. From
\eref{sigmac2}
\begin{equation}
\delta\Sigma_c(\omega)=
\left.\frac{1-t^2G_c^2(\omega)}{G_c^2(\omega)}\right|_{T=0} \delta
G_c(\omega)
\label{dsigc}
\end{equation}
and $ \delta G_c(\omega)$ can be obtained from \eref{gc_t},
\begin{equation}
\left.\left\{t^2+\left[\omega-t^2G_c(\omega)\right]
\left[\omega-3t^2G_c(\omega)\right]\right\}
\right|_{T=0} \delta G_c(\omega)= \delta \langle T^{\rm
imp}_j(\omega)\rangle^{\rm av}. 
\label{dgc}
\end{equation}
At this point it is useful to remember the definition of the T-matrix
(see \eref{tmat} and \eref{fsig})
\begin{equation}
T^{\rm imp}_j(\omega) = \frac{V^2_j}{\omega - E^f_j -
V^2_j/ \left[ \omega - t^2 G_c(\omega)\right]
 - \Sigma^{\rm imp}_{fj}\left[\omega,T,G_c\right]}.
\label{t-mat}
\end{equation}
In \eref{t-mat} we have highlighted the fact that $\Sigma^{\rm
imp}_{fj}$ is a functional of $G_c(\omega)$ as well as a function of
$\omega$ and $T$. Therefore, it is clear that the T-matrix depends on
the temperature through $\Sigma^{\rm imp}_{fj}$. However, since
$G_c(\omega)$ changes with temperature, we also need to take into
account this additional contribution, both explicitly and implicitly
through $\Sigma^{\rm imp}_{fj}$. We therefore, separate three different
contributions
\begin{eqnarray}
\delta T^{\rm imp}_j(\omega) &= \delta_1 T^{\rm imp}_j(\omega) +
\delta_2 T^{\rm imp}_j(\omega) +
\delta_3 T^{\rm imp}_j(\omega), \label{dt} \\
\delta_1 T^{\rm imp}_j(\omega) &= \left.\frac{t^2\left(T^{\rm
imp}_j(\omega)\right)^2}{\left[\omega-t^2G_c(\omega)\right]^2}\right|_{T=0}
\delta G_c(\omega), \label{dt1}\\
\delta_2 T^{\rm imp}_j(\omega) &= \left. \delta_T T^{\rm imp}_j(\omega)
\right|_{G_c^0}, 
\label{dt2} \\
\delta_3 T^{\rm imp}_j(\omega) &= \frac{\left(T^{\rm
imp}_j(\omega)\right)^2}{V^2_j} \int d\omega'
\left. \frac{\delta \Sigma^{\rm imp}_{fj}(\omega)}
{\delta G_c(\omega')}\right|_{T=0} \delta G_c(\omega').
\label{dt3}
\end{eqnarray}
In \eref{dt2}, we have used $\delta_T$ to denote the variation with $T$
which is {\em not} implicit through $G_c(\omega)$ and $G_c^0$ is used as
a reminder that the variation is calculated with a fixed
zero-temperature bath. \Eref{dt3} involves the functional derivative of
the f-self-energy with respect to the bath Green's function. Equations
\eref{dt}--\eref{dt3} can now be substituted into \eref{dgc} to yield
\begin{eqnarray}
\fl \left.\left\{t^2+\left[\omega-t^2G_c(\omega)\right]
\left[\omega-3t^2G_c(\omega)\right]
- \frac{t^2\langle\left(T^{\rm imp}_j(\omega)\right)^2\rangle^{\rm av}}
{\left[\omega-t^2G_c(\omega)\right]^2}
\right\}
\right|_{T=0} \delta G_c(\omega)  \nonumber \\
\lo- \int d\omega' \left. \langle \frac{\left(T^{\rm
imp}_j(\omega)\right)^2}{V^2_j} 
\frac{\delta \Sigma^{\rm imp}_{fj}(\omega)}
{\delta G_c(\omega')}
\rangle^{\rm av}\right|_{T=0}
 \delta G_c(\omega') 
 = \left. \langle \delta_T T^{\rm imp}_j(\omega) \rangle^{\rm av}
\right|_{G_c^0}.
\label{inteq}
\end{eqnarray}
This is an integral equation for $\delta G_c(\omega)$ whose source term
is given by the averaged variation of the T-matrix with the temperature.
Without the last term on its left-hand side, \Eref{inteq} is actually a
simple algebraic equation. The last term describes the feedback effect
that the change in the self-consistent conduction electron bath
generates on the {\em ensemble} of local impurity actions. Since raising
the temperature by a small amount leads to the unquenching of {\em a few
dilute} spins, we do not expect this feedback effect to be large and
will thus neglect the last term on the left-hand side of \eref{inteq}.

\begin{figure}
\epsfxsize=4.5in 
\centerline{\epsfbox{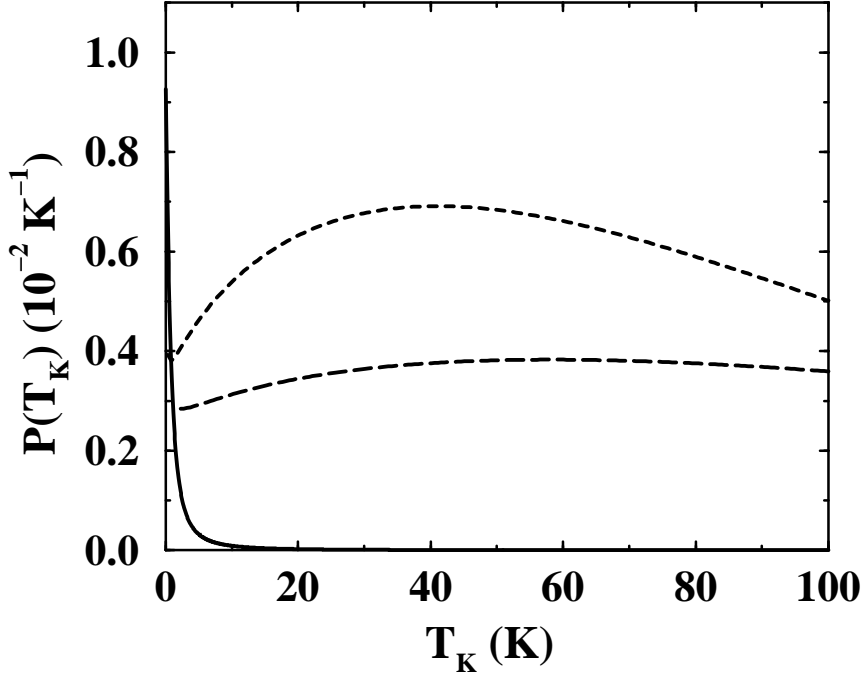}}
\caption{Experimentally determined distribution of
Kondo temperatures of the alloys UCu$_{5-x}$Pd$_x$ with
$x=1$(\longbroken) and $1.5$(\broken) (from
Ref.~\protect\cite{ucupdnmr}) and the function $F(T_K,T)$ (\full)
defined in the text.  The upturn at very low $T_K$'s is not shown. The
function $F(T_K,T)$ only probes the $T_K=0$ value of the distributions
at low $T$.
\label{fig12}}
\end{figure}

We now analyze the source term on the right-hand side of \eref{inteq}.
At zero frequency, the imaginary part of the T-matrix can be written as 
\begin{equation}
{\rm Im}T_j^{\rm imp}(T) = \frac{\sin^2\delta_{0j}}{\pi\rho_0}
t(\frac{T}{T_{Kj}}), 
\label{scalingt-mat}
\end{equation}
where  $\delta_{0j}$ is the phase shift at $T=0$. We can write down the
asymptotic limits of the scaling function $t(x)$
\begin{eqnarray}
t(x) &\approx& 1 - \alpha x^2 \qquad (x \ll 1); \nonumber \\
t(x) &\approx& \frac{\beta}{({\rm ln}(x))^2}  \qquad (x \gg 1),
\label{asympt-mat}
\end{eqnarray}
where $\alpha$ and $\beta$ are universal numbers. Therefore,
\begin{equation}
\delta_T {\rm Im} T_j^{\rm imp}(T) = -
\frac{\sin^2\delta_{0j}}{\pi\rho_0} 
\left[1 - t(\frac{T}{T_{Kj}})\right] \equiv
- \frac{\sin^2\delta_{0j}}{\pi\rho_0} 
F(T/T_{Kj}),
\label{dt-mat}
\end{equation}
which defines the function $F(x)$. This function will be averaged over
with the distribution of Kondo temperatures. Therefore, {\em keeping $T$
fixed and as as function of $T_K$}
\begin{eqnarray}
F(T/T_{Kj}) &\approx& \frac{\alpha T^2}{T_{Kj}^2} 
\qquad \qquad \quad \qquad (T_{Kj} \gg T); \nonumber \\
F(T/T_{Kj}) &\approx& 1 - \frac{\beta}{({\rm ln}(T/T_{Kj}))^2}  \qquad 
(T_{Kj} \ll T).
\label{Fasymp}
\end{eqnarray}
It can be seen that $F(T/T_{Kj})$ has a peak at $T_{Kj}=0$ with width
$T$, decaying rapidly to zero as $1/T_{Kj}^2$ for large $T_{Kj}$
(\Fref{fig12}). For low temperatures compared with the typical scale of
the distribution function $P(T_K)$ one can write
\begin{equation}
\delta_T {\rm Im} T_{\rm imp} \approx - \frac{a
\sin^2\delta_{0j}}{\pi\rho_0} T \delta(T_{Kj}),
\label{Fasdelta}
\end{equation}
where $a = \int dx F(1/x) $. A similar analysis can be carried out for
the real part of the T-matrix. Therefore, after averaging over $T_K$ one
gets
\begin{equation}
\langle \delta_T {\rm Im} T^{\rm imp}_j(\omega) \rangle^{\rm av}
\approx - \frac{a P_0
\sin^2\delta_{0}}{\pi\rho_0} T,
\label{avdt-mat}
\end{equation}
consistent with \Fref{fig11}.  It is clear that, as long as the
distribution of Kondo temperatures has finite weight at $T_K=0$, the
average T-matrix will show a linear temperature dependence. If $P(0)=0$
or negligible, then Fermi liquid behavior is recovered, with the
characteristic $T^2$ law. The result of \eref{avdt-mat} should be
plugged into \eref{inteq} and then into \eref{dsigc} for the final
expression of the conduction electron self-energy.

As our analysis above has shown, the same physical mechanism underlies
the anomalous non-Fermi liquid behavior of the resistivity as in the
case of the thermodynamic properties. It is the presence of low-$T_K$
spins, unquenched even at low temperatures, which gives rise to
anomalous scattering. Though the zero temperature transport is a
reflection of the full structure of the distribution function, the
leading low temperature behavior is much simpler, corresponding to the
gradual decrease of the number of effective scattering centers. Since
the number of released spins is small at low temperatures, their effect
is additive and an average over their subtracted T-matrices is well
justified. Finally, we add that an immediate consequence of the physical
origin of the anomalous scattering in this disorder model is a {\em
negative} magnetoresistance at low temperatures. Much like the
temperature, a magnetic field acts to destroy the low-$T_K$ Kondo
singlets and thus to suppress their effectiveness as sources of
disorder.

\section{The dynamic susceptibility}
\label{dynsusc}

The dynamic susceptibility of UCu$_{5-x}$Pd$_x$ ($x=1$ and $x=1.5$) has
been measured with inelastic neutron scattering and reported in
Ref.~\cite{ucupddyn}. The first important result was that the $\vec
q$-dependence of the magnetic response could be completely accounted for
by the $\vec q$-dependence of the Uranium ion form factor, suggesting
that the spin dynamics is completely local. Furthermore, as expected,
the frequency and temperature dependences of the imaginary part of the
dynamic susceptibility are anomalous. There was no significant
difference between the magnetic behaviors of the $x=1$ and the $x=1.5$
alloys.  These results provide a useful testing ground for the disorder
model.  Indeed, one can use the distribution of Kondo temperatures
deduced from the fits to the thermodynamic data to determine the dynamic
response and compare with the experimental data.  Like the static
magnetic susceptibility, the dynamic response will be dominated by a few
unquenched spins. Therefore, it is a reasonable assumption to take the
overall dynamic lattice response to be essentially given by an average
over the single-impurity results. Moreover, the local nature of the
measured dynamic susceptibility is consistent with this assumption.

There is currently no complete description of the dynamic susceptibility
of a single Kondo impurity for the full range of temperatures and
frequencies, though the methods to carry out this task certainly are
available. Among the existing results, we cite the unpublished work of
Costi and Hewson for the dynamic susceptibility of the Anderson model,
quoted in Ref.~\cite{costiunp}. Besides, the non-crossing approximation
(NCA) and an extension of it have also been used to determine this
response\cite{nca}. Quite often, the dynamic susceptibility of a Kondo
impurity is fitted to a relaxational form\cite{relax}
\begin{equation}
\chi''(\omega,T) = \frac{\chi(T)\Gamma(T)\omega}{\omega^2+\Gamma^2(T)},
\label{relax_dynsusc}
\end{equation}
where $\chi(T)$ is the impurity spin susceptibility and the linewidth
$\Gamma(T)$ is a function of temperature only. Note that this form
automatically satisfies the Kramers-Kronig relation
\begin{equation}
\chi(T) = \frac{1}{\pi} \int_{-\infty}^{+\infty}
\frac{\chi''(\omega,T)}{\omega} d\omega.
\label{krakro}
\end{equation}
The high temperature behavior of $\Gamma(T)$ is given the Korringa law
with logarithmic corrections\cite{walker}
\begin{equation}
\Gamma(T) \approx 4\pi\left(\rho_0J\right)^2T\left[1 - 4
\left(\rho_0J\right) {\rm ln}T\right].
\label{gammakor}
\end{equation}
At zero temperature, one can use the so-called Shiba relation to
determine $\Gamma(0)$\cite{shiba}
\begin{equation}
\lim_{\omega\to\infty} \frac{\chi''(\omega,0)}{\pi\omega} =
\frac{2\chi^2(0)}{(g\mu_B)^2} \Longrightarrow
\Gamma(0) = \frac{2T_K}{w\pi}
\label{gammashiba}
\end{equation}
where $w\approx 0.4107$ is the so-called Wilson number. In the absence
of a better description, we have employed a crude approximation that
interpolates between these two limits
\begin{equation}
\Gamma(T) = \left\{
\begin{array}{ll}
\displaystyle\frac{2T_K}{w\pi} & {\rm for}\quad T < T_K \\
4\pi\left(\rho_0J\right)^2T + \alpha & {\rm for}\quad T > T_K
\end{array}
\right. ,
\label{gammainterpol}
\end{equation}
where $\alpha$ is such that $\Gamma(T)$ is continuous. We stress the
crudeness of the approximation and regard it as a rough description of
the actual behavior. 

\begin{figure}
\epsfxsize=4.5in 
\centerline{\epsfbox{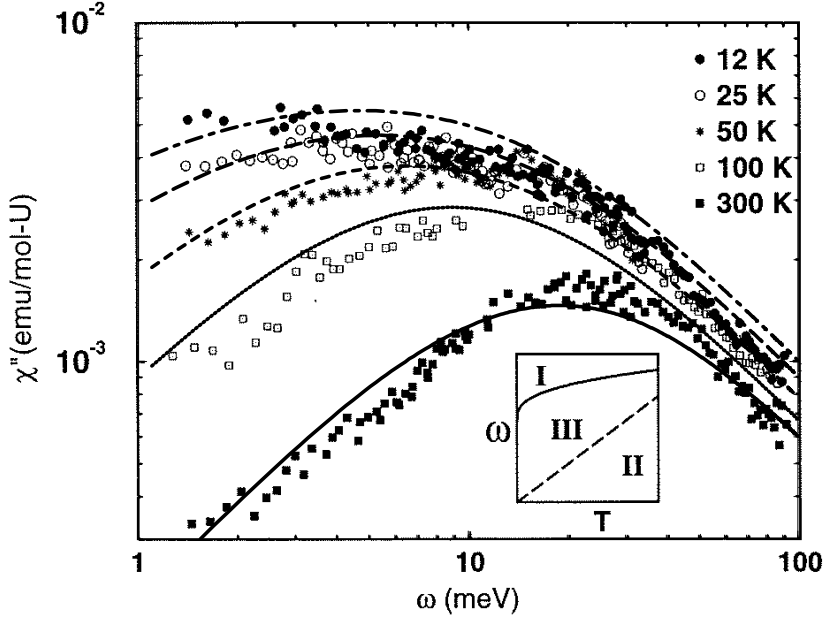}}
\caption{Comparison between the measured dynamic magnetic susceptibility
from Ref.~\protect\cite{ucupddyn} and the prediction of the disorder
model. Here, we have used the distribution function appropriate for
$x=3.5$ according to Ref.~\protect\cite{ucupdnmr}.
\label{fig13}}
\end{figure}

\begin{figure}
\epsfxsize=4.5in 
\centerline{\epsfbox{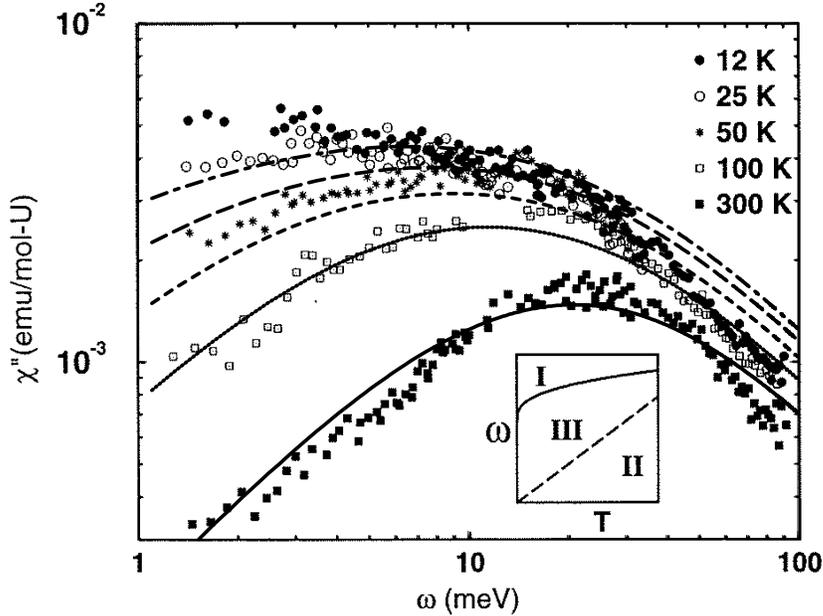}}
\caption{Comparison between the measured dynamic magnetic susceptibility
from Ref.~\protect\cite{ucupddyn} and the prediction of the disorder
model. Here, we have used the distribution function appropriate for
$x=4$ according to Ref.~\protect\cite{ucupdnmr}.
\label{fig14}}
\end{figure}

Once the behavior of the dynamic susceptibility for a Kondo impurity is
assumed to be the one given by Equations \eref{relax_dynsusc} and
\eref{gammainterpol}, one can then easily perform the average with the
distribution of coupling constants determined experimentally in
Ref.~\cite{ucupdnmr}. We have done so and the results are shown in
\Fref{fig13} ($x=3.5$) and \Fref{fig14} ($x=4$). We stress that, once
the distribution of f-shell parameters is determined from the fits to
the static magnetic susceptibility, {\em no additional fitting is
performed.} As can be seen from the figures, the agreement between the
experiment and the predictions of the disorder model is rather good,
considering the range of frequencies and temperatures and the crudeness
of our assumptions.  The agreement is slightly better when one uses the
distribution of the $x=3.5$ alloy. Though a more accurate description
should be endeavored, we believe the disorder model cannot be ruled out
by the neutron scattering data.

\section{Discussion and conclusions}
\label{discconc}

We would like to now pause and consider the overall picture that emerges
from the disorder model as well as the drawbacks of our current
treatment of the problem. We have emphasized throughout that the
important ingredient at the base of all the non-Fermi liquid features in
this approach is the presence of f-sites with arbitrarily low local
Kondo temperatures. More precisely, $P(T_K=0)\not= 0$. As the
temperature is raised, these low-$T_K$ spins are gradually unquenched
and it is this very release process that gives rise to the anomalous
thermodynamic as well as transport properties.

One of the practical difficulties one faces in applying the dynamical
field theory, is the fact that, due to the very nature of the physics
involved, one needs to be able to solve the {\em ensemble} of impurity
models in the whole range of temperatures from $T \ll T_{Kj}$ to $T \gg
T_{Kj}$. This is a notoriously difficult task as is evidenced by the
long history that led to the final solution of the Kondo problem.  The
conventional approach of Wilson's numerical renormalization group, as is
currently formulated, relies on a special ``energy-shell'' decimation
procedure, which is not obviously adapted to the present case of a
general conduction electron density of states. Therefore, in the
non-trivial case of the resistivity, we are not yet able to reliably
predict the value of the coefficient of the linear term, which might be
checked against experiments.

The dynamical mean field theory does not take into account the RKKY
interaction between f-sites. These have been studied most extensively in
the context of the two-impurity Kondo problem\cite{2impKondo}. From
these studies it is known that, if the RKKY scale is large enough
compared to $T_K$, the two impurity spins develop strong correlations
which tend to lock them into a singlet state, whereby the Kondo effect
is killed.  Since we rely on the presence of certain sites with very low
Kondo temperatures, one might wonder whether the inclusion of RKKY
interactions would not effectively provide a low energy cutoff below
which $P(T_K)$ would be essentially negligible.  Although we cannot give
a rigorous answer to this question, some very general arguments indicate
that the possibility of having low-$T_K$ spins is more robust than one
might expect.

We have already emphasized that, if the distribution function $P(T_K)$
is broad enough, at low temperatures compared to its overall width, the
fraction of spins which remain unquenched is rather small.  Therefore,
in general, they form a {\em dilute} system of spins of density
$n_{low}$, the average distance between them being proportional to
$(n_{low})^{-1/3}$. Since the average distance will be large, this will
render the RKKY interaction less effective, since its strength decays as
$R^{-3}$. Furthermore, we will argue that the effectiveness of the RKKY
interaction in suppressing the Kondo effect is actually a higher order
effect in the dilution $n_{low}$.

Indeed, consider two low-$T_K$ spins chosen at random in the sample. In
general, as argued above, they will be far apart. Now, unlike the usual
two-impurity Kondo problem, these two spins will have {\em different}
coupling constants $J_1$ and $J_2$ and, therefore, two different Kondo
scales $T_{K1}$ and $T_{K2}$. The RKKY scale $T_{RKKY} \propto J_1
J_2/D$, apart from the dependence on the distance between the spins.
Now, in the conventional two-impurity Kondo problem, where
$T_{K1}=T_{K2}$, one has, qualitatively, two independent Kondo screening
processes, one at each spin, when $T_{K1}=T_{K2} \gg T_{RKKY}$, whereas
the effect will be suppressed when $T_{K1}=T_{K2} \ll T_{RKKY}$. In our
disordered case, RKKY dominates if, say, $T_{K1} < T_{K2} \ll T_{RKKY}$.
However, in the intermediate case when $T_{K1} \ll T_{RKKY} \ll T_{K2}$,
two independent Kondo effects will survive, even though $T_{RKKY}$ is
still larger than one of the Kondo temperatures. Indeed, as one comes
down in energy scale, by the time one hits $T_{RKKY}$, spin $2$ has
already undergone quenching and is no longer available to correlate into
a singlet-like composite with spin $1$. Therefore, the RKKY
effectiveness depends on the random selection of two spins, both of
which must have low enough Kondo temperatures. We would thus expect it
to be of order $n_{low}^2$. Naturally, the transitions between different
regimes are all crossovers (in the absence of special unphysical
symmetries) and there will be no sharp distinctions between the
different cases and hence no sharp cutoff to the distribution function
$P(T_K)$.

We have emphasized throughout our analysis the importance of the
structure of the distribution of Kondo temperatures, in particular,
whether it intercepts the vertical axis at $T_K=0$ or not. Due to the
exponential dependence of $T_K$ on the f-parameters, very small
modifications in the width of the bare distribution lead to rather large
changes in the distribution of Kondo temperatures. If the bare width is
too small, $P(T_K)$ will be negligible at low $T_K$, whereas, if it is
too large, $P(T_K)$ will be divergent. Therefore, the situation where
$P(T_K)$ can be taken as approximately constant when $T_K \to 0$ only
holds in a narrow window of bare disorder widths, in our current
approach. In a more speculative vein, one is tempted to think that the
RKKY interaction, which is left out of the current approach, might
intervene to impose an upper bound on the density of low-$T_K$ spins
which are allowed to exist in such a state. It would do so by
effectively eliminating them through the formation of random
singlets\cite{localdis1}. 

Finally, by treating the disorder seen by the conduction electrons in a
mean-field CPA-like fashion, we have neglected the effects of
fluctuations in the conduction electron density of states. It is natural
to expect that these effects might act to further renormalize the
distribution of Kondo temperatures. It would be interesting to study
whether, like in many other treatments of disordered
systems\cite{localdis1}, these effects lead to a flow towards a
universal form of the distribution function.

In conclusion, we have presented a complete picture of the possible
origin of the non-Fermi liquid behavior in Kondo alloys. Our study has
shown the important effect correlations have on the role of disorder in
f-electron systems, considerably enhancing the bare f-shell disorder
strength. The interplay between disorder and correlations leads to the
idea of a distribution of Kondo temperatures, whose structure determines
whether a Fermi liquid description is possible. In the particular case
when $P(0)\not=0$, both thermodynamic and transport properties are
anomalous and incompatible with a Fermi liquid picture.

\ack

We would like to acknowledge particularly helpful and insightful
comments by D. MacLaughlin.  We are also grateful to B. Andraka, M. C.
Aronson, N.  Bonesteel, A. H. Castro Neto and J. R.  Schrieffer for
useful discussions.  This work was supported by the National High
Magnetic Field Laboratory at Florida State University. GK was supported
by NSF DMR 95-29138.

\appendix

\section{Simplified expression for the conductivity}
\label{app_expcond}

In the case of a semicircular density of states, a simplification of the
expression \eref{dccond} can be achieved. Indeed\cite{infd_andlat_b},
\begin{eqnarray}
I &= \int d\epsilon \rho_0(\epsilon) A^2(\epsilon,\omega) = 
\int d\epsilon \rho_0(\epsilon) \frac{1}{2} {\rm Re} \left[
G_c(\epsilon,\omega) G_c^*(\epsilon,\omega) - G_c^2(\epsilon,\omega)
\right] \nonumber \\
&= \frac{1}{2} \left\{ {\rm Re} \left[ \frac{\partial G_l(z)}{\partial
z}\right] - \frac{{\rm Im} G_l(z)}{{\rm Im} z} \right\},
\label{simpl}
\end{eqnarray}
where $z \equiv \omega + \mu - \Sigma_c(\omega)$ and 
\begin{equation}
G_l(z) \equiv \int d\epsilon \frac{\rho_0(\epsilon)}{z - \epsilon}.
\label{glocal}
\end{equation}
Using the identity $z = G_l^{-1}(z) + t^2 G_l(z)$, which follows from
\eref{sigmac}, we get
\begin{equation}
\fl \sigma_{DC} = \frac{2e^2}{\hbar\pi a} \int_{-\infty}^{+\infty}
d\omega \left( - \frac{\partial f}{\partial \omega}\right)
\left\{ {\rm Re} \left[ \frac{1}{(t^2 G_l(z))^2 -1} \right]
+ \frac{1}{1 - |t^2 G_l(z)|^2} \right\}.
\label{dccondsimpl}
\end{equation}

\section{Conduction electron self-energy and the impurity model
T-matrix} 
\label{app_cself_tmat}

The expression for the conduction electron self-energy
$\Sigma_c(\omega)$ can be recast in a more illuminating form by
employing the T-matrix associated with the impurity problem, which is
defined by
\begin{equation}
T^{\rm imp}_j(\omega) \equiv V^2_j G^{\rm imp}_{fj}(\omega).
\label{tmat}
\end{equation}
Inserting this definition into \eref{fsig}, then into \eref{phi} and
\eref{gc}, we can write after some manipulations
\begin{eqnarray}
\overline{G}_c(\omega) &=& \frac{1}{\omega + \mu - t^2
\overline{G}_c(\omega)} 
+ \frac{\langle T^{\rm imp}_j(\omega)\rangle^{\rm av}}{(\omega + \mu -
t^2 \overline{G}_c(\omega))^2}.
\label{gc_t}
\end{eqnarray}
Now, \eref{sigmac} can be explicitly solved for a semicircular density
of states giving
\begin{eqnarray}
\Sigma_c(\omega) &=& \omega + \mu - \frac{1}{\overline{G}_c(\omega)} -
t^2 \overline{G}_c(\omega).
\label{sigmac2}
\end{eqnarray}
Combining \eref{gc_t} and \eref{sigmac2} we finally get
\begin{eqnarray}
\Sigma_c(\omega) &=& \frac{\langle T^{\rm imp}_j(\omega)\rangle^{\rm
av}}{\overline{G}_c(\omega)(\omega + \mu - t^2 \overline{G}_c(\omega))}.
\label{sigmac3}
\end{eqnarray}

\section{Thermodynamic properties}
\label{thermo}

In this appendix, we will investigate whether the assumptions of the
simple disorder model of Ref.~\cite{ucupdnmr} (hereafter, SDM) are
justified within the framework of the dynamical mean field theory. Let
us recapitulate what the assumptions of the SDM are.  Essentially, the
model assumes a collection of independent Anderson or Kondo impurities,
each of which hybridizes with the same, featureless conduction bath. The
parameters of this collection of impurities are distributed according to
some assumed distribution function(s). In the particular case of
Ref.~\cite{ucupdnmr}, which assumed Kondo spins, the dimensionless Kondo
coupling constant $\lambda \equiv \rho_0 J$ was assumed to be
distributed according to a Gaussian. Thermodynamic properties are then
calculated by taking the average over the single-impurity results with
the appropriate distribution functions, as if the impurity responses
were completely uncorrelated.

It is clear that the dynamical mean field theory retains much of the
flavor of the SDM, with the {\em ensemble} of impurity problems playing
the role of the uncorrelated collection of spins. The question we would
like to answer is whether the equivalence can be shown within the
framework of the dynamical mean field theory and, more importantly, what
are the conditions for the validity of the equivalence. We will confine
our analysis to the total energy, which is enough for the calculation of
specific heat. The same conclusions apply to the susceptibility, which,
however, requires consideration of the free energy.

Let us first write down the total energy in the framework of the
dynamical mean field theory. We have
\begin{eqnarray}
\fl E_{DMF} = \langle H \rangle = \sum\limits_{ij\sigma} - t_{ij} \left( 
\langle c^{\dagger}_{i\sigma} c^{\phantom{\dagger}}_{j\sigma}\rangle +
{\rm H.c.} \right)
+ \sum\limits_{j\sigma} E^f_{j} \langle f^{\dagger}_{j\sigma}
f^{\phantom{\dagger}}_{j\sigma}\rangle \nonumber \\
+ \sum\limits_{j\sigma} V_j \left( \langle c^{\dagger}_{j\sigma}
f^{\phantom{\dagger}}_{j\sigma} \rangle + {\rm H. c.} \right)
+ U \sum\limits_{j} \langle n_{fj\uparrow} n_{fj\downarrow}\rangle.
\label{mfenergy}
\end{eqnarray} 
We can write
\begin{equation}
\langle c^{\dagger}_{i\sigma} c^{\phantom{\dagger}}_{j\sigma}\rangle =
T \sum_{i\omega_n} G_{c\sigma}^{ij}(i\omega_n).
\label{hopping}
\end{equation}
Now, when the coordination number $z$ goes to infinity and for the case
of a Bethe lattice (in the absence of symmetry breaking) one can prove
that\cite{lisareview} 
\begin{equation}
G_{c\sigma}^{ij}(i\omega_n) = - t_{ij} G_c^2(i\omega_n),
\label{hopping2}
\end{equation}
from which it follows
\begin{equation}
\fl \sum_{ij\sigma} - t_{ij} \left( 
\langle c^{\dagger}_{i\sigma} c^{\phantom{\dagger}}_{j\sigma}\rangle +
{\rm H.c.} \right) = 2T \sum_{i\omega_n}\sum_{ij} t^2_{ij}
G_c^2(i\omega_n) = 2 {\cal N} T t^2 \sum_{i\omega_n}
\overline{G}_c^2(i\omega_n),
\label{hopping3}
\end{equation}
where $\cal N$ is the number of lattice sites and, consistent with the
correct rescaling of the hopping in the infinite coordination limit,
$t^2 \equiv z t^2_{ij}$. Plugging \eref{hopping3} into \eref{mfenergy},
one then has the total energy completely expressed in terms of local
quantities. This is the quantity we want to compare with the prediction
of the SDM. 

In order to do that, one needs to define what precisely one means by the
SDM. In the dynamical mean field theory, we have the {\em ensemble} of
impurity models defined by \Eref{effaction}. We, therefore, {\em define}
the total energy in the SDM by the sum of the energies of the various
impurities
\begin{equation}
E_{SDM} = \sum_j E^{\rm imp} \left\{E^f_j,V_j\right\}.
\label{sdmenergy}
\end{equation}
To define the energy of one impurity, we first write the Hamiltonian
corresponding to \Eref{effaction}
\begin{equation}
\fl H^{\rm imp} \left\{E^f,V\right\} = \sum_{\vec k\sigma}
E_{\vec k} 
a^{\dagger}_{\vec{k}\sigma} a^{\phantom{\dagger}}_{\vec{k}\sigma}
+ \sum_{\sigma} E^f f^{\dagger}_{\sigma}
f^{\phantom{\dagger}}_{\sigma} 
+ V \sum_{\vec k\sigma} \left(  a^{\dagger}_{\vec{k}\sigma}
f^{\phantom{\dagger}}_{\sigma} + {\rm H.c.} \right) 
+ U \sum_{j} n_{f\uparrow} n_{f\downarrow},
\label{imphammy}
\end{equation}
where we introduced fictitious fermionic operators $a_{\vec{k}\sigma}$
to mimic the self-consistent conduction electron bath through the
``cavity'' Green's function of \Eref{delta}
\begin{equation}
\Delta(\omega) \equiv \sum_{\vec k} \frac{1}{\omega - E_{\vec k}}.
\label{cavity}
\end{equation}
We then define the appropriate single-impurity energy by
\begin{equation}
E^{\rm imp} \left\{E^f_j,V_j\right\} \equiv \left\langle H^{\rm imp}
\left\{E^f,V\right\} \right\rangle - \sum_{\vec k\sigma} \left. E_{\vec
k}  \langle a^{\dagger}_{\vec{k}\sigma}
a^{\phantom{\dagger}}_{\vec{k}\sigma}\rangle\right|_{V=0},
\label{impenergy}
\end{equation}
where we subtracted the total energy of the fictitious conduction
electron bath evaluated at $V=0$. Now, the fictitious conduction
electron Green's function is
\begin{equation}
G_a(\vec k, \vec k',\omega) = \delta_{\vec k \vec k'} G^{0}_a(\vec
k,\omega) + G^{0}_a(\vec k,\omega) V^2 G^f(\omega) G^{0}_a(\vec
k',\omega).
\label{ficgc}
\end{equation}
Using
\begin{equation}
\langle a^{\dagger}_{\vec{k}\sigma}
a^{\phantom{\dagger}}_{\vec{k}\sigma} \rangle = T \sum_{i\omega_n}
G_a(\vec k, \vec k,i\omega_n),
\label{adaga}
\end{equation}
we have
\begin{equation} \langle a^{\dagger}_{\vec{k}\sigma}
a^{\phantom{\dagger}}_{\vec{k}\sigma}\rangle -
\left. \langle a^{\dagger}_{\vec{k}\sigma}
a^{\phantom{\dagger}}_{\vec{k}\sigma}\rangle\right|_{V=0} =
T V^2 \sum_{i\omega_n} G^f(i\omega_n)
\left[ G^{0}_a(\vec k,i\omega_n)\right]^2.
\label{deladaga}
\end{equation}
Thus
\begin{eqnarray}
\fl \sum_{\vec k\sigma} E_{\vec k} \left[ 
\langle a^{\dagger}_{\vec{k}\sigma}
a^{\phantom{\dagger}}_{\vec{k}\sigma}\rangle -
\left. \langle a^{\dagger}_{\vec{k}\sigma}
a^{\phantom{\dagger}}_{\vec{k}\sigma}\rangle\right|_{V=0}
\right]  =
2T V^2 \sum_{i\omega_n} G^f(i\omega_n) \sum_{\vec k} \frac{E_{\vec
k}}{\left(i\omega_n - E_{\vec k}\right)^2}  \nonumber \\
= -2T V^2 \sum_{i\omega_n} G^f(i\omega_n) \left[
\Delta(i\omega_n)  + i\omega_n
\frac{\partial\Delta(i\omega_n)}{\partial(i\omega_n)}\right].
\label{kinetic}
\end{eqnarray}
In the last equality we used \Eref{cavity}.

We can now use Equations \eref{mfenergy}, \eref{sdmenergy},
\eref{imphammy} and \eref{impenergy} to write the difference in energy
between the SDM and the dynamical mean field theory
\begin{equation}
\fl \Delta E \equiv E_{DMF} - E_{SDM} = \sum_{\vec k\sigma}
\left\{E_{\vec k}  
\left[\left. \langle a^{\dagger}_{\vec{k}\sigma}
a^{\phantom{\dagger}}_{\vec{k}\sigma}\rangle\right|_{V=0} -
\langle a^{\dagger}_{\vec{k}\sigma}
a^{\phantom{\dagger}}_{\vec{k}\sigma}\rangle \right] +
\left(\epsilon_{\vec k} - \mu\right) \langle 
c^{\dagger}_{\vec{k}\sigma} 
c^{\phantom{\dagger}}_{\vec{k}\sigma} \rangle \right\}.
\label{delenergy}
\end{equation}
To simplify things further, we can use Equations \eref{hopping3} and
\eref{kinetic} to get
\begin{eqnarray}
\fl \Delta E =  2T\sum_{i\omega_n}\left\{
{\cal N}  t^2 \overline{G}_c^2(i\omega_n) +
\sum_j V^2_j G^f_j(i\omega_n) \left[
\Delta(i\omega_n) + i\omega_n
\frac{\partial\Delta(i\omega_n)}{\partial(i\omega_n)}\right]\right\}.
\label{delenergy2}
\end{eqnarray}
We then use \Eref{gc_t} and the definition \eref{delta} to arrive at our
final expression for the difference in energy between the dynamical mean
field theory and the SDM
\begin{eqnarray}
\frac{\Delta E}{\cal N} = 2T\sum_{i\omega_n}\left\{
t^2 \overline{G}_c^2(i\omega_n) +
\left[\frac{\overline{G}_c(i\omega_n)}{\Delta(i\omega_n)} - 1 \right]
\left[1 +
\frac{\partial{\rm ln}\Delta(i\omega_n)}
{\partial{\rm ln}(i\omega_n)}\right]\right\}. 
\label{delenergy3}
\end{eqnarray}

We are now in a position to determine whether the SDM is accurate enough
to give the thermodynamic properties of the system within the dynamical
mean field theory framework. For this, it is enough to consider
\Eref{delenergy}. All the quantities in this equation are related to
conduction electron kinetic properties. The corresponding densities of
states are given by the imaginary parts of either
$\overline{G}_c(\omega)$ or the ``cavity'' Green's function
$\Delta(\omega)$. As can be clearly seen from Figures \ref{fig3} and
\ref{fig4}, these quantities are weakly renormalized. Their
contributions to, say, the specific heat coefficient $\gamma = C_V(T)/T$
are of the order of $1/D$ and are completely negligible when compared to
the contribution from the impurity part, which is of order $1/T_K \gg
1/D$. Therefore, as far as thermodynamic properties such as $\gamma$ and
$\chi$ are concerned, the approximation of averaging over
single-impurity results is perfectly consistent with the solution of the
full dynamical mean field theory. This is true of both the clean and the
dirty systems, since nothing in this argument relied on the presence of
disorder. However, we stress the fact that the {\em ensemble} of
impurity problems must be solved in the {\em fully self-consistent
conduction electron bath}. Though the renormalizations of this bath are
small, they affect the impurity properties through the conduction
electron density of states $\rho_0$. The latter quantity appears in the
argument of the exponential in the expression for the Kondo temperature
and can, therefore, lead to rather substantial renormalizations of the
total energy. To fully describe these changes, one needs to solve the
full dynamical mean field theory equations.

\section{The slave boson mean field theory of the impurity models}
\label{app_slbos}

The slave boson description of the infinite-U Anderson Model starts with
the replacement of the non-holonomic constraint $n_f < 1$ imposed by the
$U \rightarrow \infty$ condition by a new set of slave boson operators
$b_j$ together with a holonomic constraint\cite{slavebos}
\begin{eqnarray}
f_{j\sigma} &\longrightarrow b_j^{\dagger}f_{j\sigma} \label{slsubs_a}
\\ n_{fj\sigma} &\longrightarrow n_{fj\sigma} \label{slsubs_b} \\
\sum_{\sigma} \left( n_{fj\sigma} \right) < 1 &\longrightarrow
\sum_{\sigma} \left( n_{fj\sigma} \right) + b_j^{\dagger}b_j  =1.
\label{slsubs_c}
\end{eqnarray}
The constraint is then imposed by introducing a Lagrange multiplier term
in the Lagrangian
\begin{equation}
{\cal L}_0 \longrightarrow {\cal L}_0 + i \int_0^{\beta} d\tau \lambda_j(\tau)
\left[ \sum_{\sigma} 
f^{\dagger}_{fj\sigma}(\tau)f^{\phantom{\dagger}}_{fj\sigma}(\tau)
  + b_j^{\dagger}(\tau)b_j(\tau) -1
\right] ,
\label{lagr}
\end{equation}
where $\lambda_j(\tau)$ is an additional bosonic field variable. The
transformed action corresponding to the impurity action \eref{effaction}
is then 
\begin{eqnarray}
\fl S^{\rm imp}_j &= S^{\rm imp}_{1j} + S^{\rm imp}_{2j}; \nonumber \\
\fl S^{\rm imp}_{1j} &=
\int_0^{\beta} d\tau \left[
\sum_{\sigma} f^{\dagger}_{j\sigma}(\tau) \left[
\partial_{\tau} + E^f_j + i \lambda_j(\tau) \right] 
f^{\phantom{\dagger}}_{j\sigma}(\tau) 
+ b^{\dagger}_{j}(\tau) \left[ \partial_{\tau} + i \lambda_j(\tau) \right]
b^{\phantom{\dagger}}_{j}(\tau) - i \lambda_j(\tau)
\right]; \nonumber \\
\fl S^{\rm imp}_{2j} &=
\int_0^{\beta} d\tau \int_0^{\beta} d\tau '
\sum_{\sigma} f^{\dagger}_{j\sigma}(\tau)
b^{\phantom{\dagger}}_{j}(\tau) 
V^2_j\Delta(\tau - \tau ')
f^{\phantom{\dagger}}_{j\sigma}(\tau ')
b^{\dagger}_{j}(\tau '),
\label{lagrangian}
\end{eqnarray}
where $\Delta(\tau)$ is the Matsubara--Fourier transform of
\eref{delta}. Note that the second part of the impurity action, which
corresponds to hybridization processes between the f-site and the
effective conduction electron bath, has been modified by the
introduction of the slave boson field $b_j(\tau)$, which makes the
bookkeeping of the occupation of the f-site.

At the mean field level, the bosonic fields $b_j(\tau)$ and
$\lambda_j(\tau)$ acquire a time-independent expectation value and behave
as c-numbers\cite{slavebos}. The resulting effective action is quadratic
in the pseudo f-electrons and can be exactly solved. Following the
convention of writing $i \langle \lambda_j \rangle = \epsilon^f_j -
E^f_j$ and $\langle b_j \rangle = r_j$, it reads
\begin{equation}
\fl S^{\rm eff}_j = T \sum_{\omega_n\sigma} \left[
f^{\dagger}_{j\sigma}(i\omega_n)
\left( - 
i\omega_n + \epsilon^f_j + r^2_j V^2_j \Delta(i\omega_n) \right)
f^{\phantom{{\dagger}}}_{j\sigma}(i\omega_n) 
\right] + \left( \epsilon^f_j - E^f_j \right) \left( r^2_j-1 \right).
\label{effaction_sb}
\end{equation}
The mean field parameters $r_j$ and $\epsilon^f_j$ are determined by a
saddle point extremization of the free energy corresponding to
\eref{effaction_sb}. The mean field equations then read
\begin{eqnarray}
\frac{2}{\pi} \int_{-\infty}^{\infty} d\omega
\frac{f(\omega)\tilde{\Delta}_j''(\omega)}
{\left(\omega - \epsilon^f_j - \tilde{\Delta}_j'(\omega)\right)^2 +
\left(\tilde{\Delta}_j''(\omega)\right)^2}
+ r_j^2 - 1 &= 0; \label{mfeqsa}\\
\frac{2}{\pi} \int_{-\infty}^{\infty} d\omega
\frac{f(\omega)\tilde{\Delta}_j''(\omega)\left(\omega -
\epsilon^f_j\right)}
{\left(\omega - \epsilon^f_j - \tilde{\Delta}_j'(\omega)\right)^2 +
\left(\tilde{\Delta}_j''(\omega)\right)^2}
+ r_j^2 \left(\epsilon^f_j - E^f_j\right) &= 0,
\label{mfeqsb}
\end{eqnarray}
where $f(\omega)$ is the Fermi function, $\tilde{\Delta}_j(\omega)=r^2_j
V^2_j \Delta(\omega)$ and single and double primes denote real and
imaginary parts, respectively. We note that, given the hybridization
function $\Delta(i\omega_n)$, each value of the bare parameters $E^f_j$
and $V_j$ will define a different impurity problem with its own
corresponding values of $r_j$ and $\epsilon^f_j$. Thus, a distribution
of mean field parameters is also generated.

The mean field treatment we have described gives the following
expression for the f-electron Green's function
\begin{equation}
G_{fj}(i\omega_n) = \frac{r^2_j}{i\omega_n - \epsilon^f_j - r^2_j
V^2_j \Delta(i\omega_n)}.
\label{fgreen_sb}
\end{equation}
Note that the numerator in \eref{fgreen_sb} is crucial. It is a
consequence of the slave boson prescription \eref{slsubs_a} and
distinguishes the pseudo f-electron Green's function, from which it is
absent, from the real f-electron Green's function. Finally, using the
definition \eref{fsig}, one can write the mean field expression for the
f-electron self-energy
\begin{equation}
\Sigma_{fj}(i\omega_n) = i\omega_n -E^f_j -
\frac{\left(i\omega_n - \epsilon^f_j
\right)}{r^2_j},
\label{fsig_sb}
\end{equation}
which yields
\begin{equation}
\Phi_j(i\omega_n) = \frac{r^2_j V^2_j}{i\omega_n - \epsilon^f_j}.
\label{phi_sb}
\end{equation}
This can then be inserted into \eref{gc}, thus closing the
self-consistency loop. Note that, in the pure case,
\begin{equation}
\Sigma_c(\omega) = \Phi(\omega) = \frac{r^2 V^2}{\omega -
\epsilon^f_j},
\label{pure_sb}
\end{equation}
which is consistent with the low-energy Fermi liquid parameterization of
\eref{csig_fl}.

\end{document}